\documentclass[aps,prb,twocolumn,noeprint,superscriptaddress]{revtex4-1}

\usepackage[log-declarations=false]{xparse}
\usepackage[pdftex]{graphicx}
\usepackage{physics}
\usepackage{amssymb,amsmath}
\usepackage{subscript}
\usepackage{xcolor}
\usepackage{color}
\usepackage{hyperref}

\hypersetup{colorlinks=false,urlcolor=black}
\newcommand{\mbf}[1]{\ensuremath{\mathbf{#1}}}
\newcommand{\bb}{\mbf{b}}

\begin{document}

\title{Learning electron densities in the condensed phase}

\author{Alan M.~Lewis}
\email{alan.lewis@mpsd.mpg.de}
\affiliation{Max Planck Institute for the Structure and Dynamics of Matter, Luruper Chaussee 149, 22761 Hamburg, Germany}

\author{Andrea Grisafi}
\affiliation{Laboratory of Computational Science and Modeling, IMX, \'Ecole Polytechnique F\'ed\'erale de Lausanne, 1015 Lausanne, Switzerland}

\author{Michele Ceriotti}
\affiliation{Laboratory of Computational Science and Modeling, IMX, \'Ecole Polytechnique F\'ed\'erale de Lausanne, 1015 Lausanne, Switzerland}

\author{Mariana Rossi}
\email{mariana.rossi@mpsd.mpg.de}
\affiliation{Max Planck Institute for the Structure and Dynamics of Matter, Luruper Chaussee 149, 22761 Hamburg, Germany}

\begin{abstract}
We introduce a local machine-learning method for predicting the electron densities of periodic systems. The framework is based on a numerical, atom-centred auxiliary basis, which enables an accurate expansion of the all-electron density in a form suitable for learning isolated and periodic systems alike. We show that using this formulation the electron densities of metals, semiconductors and molecular crystals can all be accurately predicted using symmetry-adapted Gaussian process regression models, properly adjusted for the non-orthogonal nature of the basis. These predicted densities enable the efficient calculation of electronic properties which present errors on the order of tens of meV/atom when compared to \emph{ab initio} density-functional calculations. We demonstrate the key power of this approach by using a model trained on ice unit cells containing only 4 water molecules to predict the electron densities of cells containing up to 512 molecules, and see no increase in the magnitude of the errors of derived electronic properties when increasing the system size. Indeed, we find that these extrapolated derived energies are more accurate than those predicted using a direct machine-learning model. Finally, on heterogeneous datasets SALTED can predict electron densities with errors below 4\%.
\end{abstract}

\maketitle

\section{Introduction}

The electron density $\rho$ is a fundamental quantity of quantum chemistry and physics, which in principle can determine all of the ground state properties of a system. Using density-functional theory (DFT), a wide variety of these properties can be derived directly from the electron density, such as energies, charges, dipoles, and electrostatic potentials.\cite{Cohen2000,Sousa2007,Cohen2012,Koch2015} As a result, obtaining accurate electron densities is central to many applications within computational chemistry, physics and material science. 

In DFT, the ground state electron density is found by performing a constrained minimization of the energy functional.\cite{Levy1979} This is most commonly achieved by self-consistently solving the Kohn-Sham equations.\cite{Hohenberg1964,Kohn1965} This minimization procedure is expensive, and formally scales with the cube of the number of electrons in the system,\cite{Sousa2007} although implementations which approach linear scaling are available.\cite{Blum2009,Prentice2020,Nakata2020} As a result, while DFT computations are hugely successful and widely used, they remain limited by the system size: typically they can be applied to at most a few thousand atoms.\cite{Ratcliff2017} Furthermore, when DFT is used to perform \emph{ab initio} molecular dynamics, many successive DFT calculations are required on very similar structures, severely limiting the time-scales available to these simulations.

In recent years, methods have been proposed which use machine-learning techniques to predict electron densities while avoiding the need to minimize the energy functional. For example, Alred and coworkers have reported a method to directly predict the electron density on a real-space grid, where each grid-point is used to provide a local representation of the atomic structure,\cite{Alred2018} a strategy that was also followed by Chandrasekaran et al.\cite{Chandrasekaran2019} However, the sheer number of grid points which must be used to accurately represent the density in this way significantly increases the computational cost of this approach. Limiting the dimensionality of the learning problem can be achieved by representing the scalar field using a finite number of basis functions. Brockherde et al. introduced a framework that makes use of a plane-wave basis and constructed a separate kernel-based model to regress each individual Fourier component of the pseudo-valence electron density.\cite{Brockherde2017,bogo+20nc} While the choice of plane waves carries the advantage of allowing a systematic convergence of the scalar field in the limit of an infinitely large basis, adopting a set of center-less functions to discretize the learning problem limits the application of the method to relatively rigid systems which can be unambiguously aligned along a prescribed orientation.

To overcome these hurdles, a method capable of predicting the electron density of a system when represented using an atom-centered spherical harmonic basis was recently introduced.\cite{Grisafi2019} The problem is recast as the regression of a set of local density coefficients which can be predicted in a rotationally-covariant fashion thanks to the \textit{symmetry-adapted Gaussian process regression} (SA-GPR) framework.~\cite{gris+18prl} The SA-GPR framework has been used in several contexts,\cite{Raimbault:2019,wilk+19pnas,Yang:2019} and most importantly enabled the data-efficient and highly-transferable learning of the electron density for arbitrarily complex molecular systems.\cite{gris+18prl} In a follow-up work by Corminbeouf and coworkers, the local expansion of the density field has been made coherent with state-of-the-art \textit{resolution of the identity} (RI) schemes, giving access to reference electron densities which show an accuracy comparable to that of common quantum-chemical calculations.\cite{Fabrizio2019,fabr+20chimia} 

In this paper we extend the method of Refs.~\citenum{Grisafi2019} and \citenum{Fabrizio2019} to the condensed phase and demonstrate its applicability to a wide variety of systems. By using numerical atom-centred basis functions first developed for use in RI schemes of the exchange operator,\cite{Ren2012} we retain the local nature of the method, and can treat periodic and finite molecular systems on the same footing. We demonstrate that for a range of test systems expanding the density in this basis introduces only small and controllable errors in both the density itself, and energies derived from the density.

We will refer to the application of SA-GPR within a machine-learning model that is capable of performing the regression of electron densities both in finite and periodic systems as the \textit{symmetry-adapted learning of three-dimensional electron densities} (SALTED) method. We employ SALTED to produce a series of regression models which are applied to predict the electron density of a metal, a semiconductor and a molecular solid in turn, obtaining for each of these systems accurate densities with fewer than 100 training structures, as well as derived electronic properties that present errors on the order of tens of meV/atom. Finally, we use a model trained on ice cells containing four molecules to predict the densities and derived energies of up to 512-molecule supercells. We see no loss of accuracy in these energies as the size of the target system increases, indicating that our local learning framework is sufficiently transferable to capture the information needed to predict the energy of extended systems. Furthermore, we find that these extrapolated derived energies are more accurate than those predicted using a direct machine-learning model.

\section{Theory}
\label{sec:theory}

\subsection{The RI framework}

The periodic electron density $\rho(\mathbf{r})$ may be expanded as a linear combination of atom-centred basis functions using a resolution of the identity (RI) ansatz:
\begin{equation}
\begin{aligned}
\rho(\mathbf{r}) \approx \tilde{\rho}(\mathbf{r}) & = \sum_{i,\sigma,\mathbf{U}} c_{i,\sigma} \phi_{i,\sigma}(\mathbf{r} - \mathbf{R}_i + \mathbf{T}(\mathbf{U})) \\ 
& = \sum_{i,\sigma,\mathbf{U}} c_{i,\sigma} \braket{\mathbf{r}}{\phi_{i,\sigma}(\mathbf{U})} .
\end{aligned}
\label{approx_rho}
\end{equation}
Here $\mathbf{R}_i$ is the position of atom $i$, the basis function $\phi_{i,\sigma}$ is centered on atom $i$ and may be written as the product of a radial part $R_n(r)$ and a real spherical harmonic $Y_{\lambda\mu}(\theta,\phi)$, so that we make use of composite index $\sigma \equiv (n\lambda\mu)$. $\mathbf{T}(\mathbf{U})$ is a translation vector to a point removed from the central reference unit cell by an integer multiple $\mathbf{U} = (U_x,\, U_y,\, U_z)$ of the lattice vectors. The expansion coefficients $c_{i,\sigma}$ then completely define the approximate density $\tilde{\rho}(\mathbf{r})$. This RI ansatz has long been used in effective single-particle approximations of the electronic energy, such as Hartree-Fock, M{\o}ller-Plesset and Kohn-Sham DFT, with the purpose of bypassing the unfavourable scaling of computing the 4-electron-2-center integrals that underlie the definition of the Hartree energy,~\cite{bearends1973,weigend+06pccp,golze+17jctc} as well as of the exact exchange introduced in hybrid exchange-correlation functionals.\cite{sodt+08jcp,manzer+15jctc}

\begin{figure}[t]
    \centering
    \includegraphics[width=8cm]{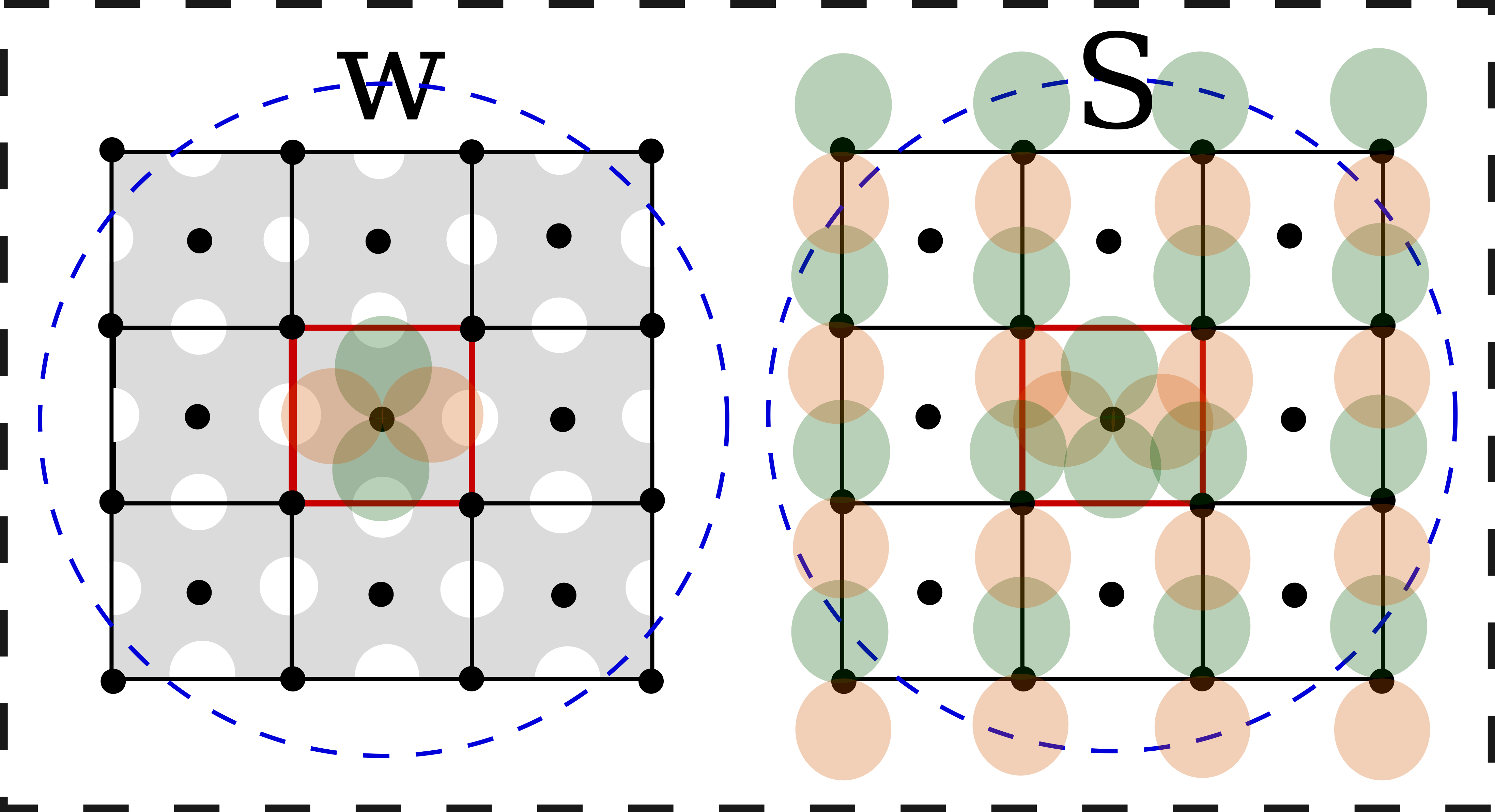}
    \caption{(\textit{left}) The unfolded representation of the projection $\mathbf{w}$ of the periodic electron density (shaded grey area) onto atom-centered basis functions. (\textit{right}) The unfolded representation of the periodic overlap $\mathbf{S}$ between atom-centered basis functions. The central red cell indicates the special unit cell used to compute the periodic integrals in the folded representation (see text). }
    \label{fig:method}
\end{figure}

Different error metrics can be adopted to determine the expansion coefficients, which influence the accuracy that one is willing to achieve on prescribed classes of density-derived properties.~\cite{briling+21arxiv} A Coulomb metric, for instance, is typically used to provide RI approximations that give minimal error in the Hartree energy.~\cite{whitten1973jcp} In this work, we define the RI expansion coefficients as those which minimize the integral over a single unit cell of the square error in the density itself, i.e.,
\begin{equation}
\epsilon(\mathbf{c}^{\text{RI}}) = \int_{u.c.} d\mathbf{r} \left| \tilde{\rho}^{\text{RI}}(\mathbf{r};\mathbf{c}^{\text{RI}}) - \rho^{\text{QM}}(\mathbf{r}) \right|^2,
\label{eq:loss_func1}
\end{equation}
where $\rho^{\text{QM}}(\mathbf{r})$ is the self-consistent electron density which we are using as the fitting target. Note that we do not impose any constraint on the conservation of the number of electrons when calculating the RI coefficients. In fact, we find that including this constraint results in undue weight being given to the isotropic basis functions ($\lambda=0$), relative to an unconstrained minimization, damaging the accuracy of the overall scalar-field representation. Moreover, as clarified in Ref.~\citenum{briling+21arxiv}, imposing a corresponding constraint on the machine-learning model would only limit the electronic charge conservation to those structures that are used for training, and including such a constraint inevitably leads to a breakdown in the stability of the machine-learning model as the number of training structures is increased.\cite{briling+21arxiv}

Minimization of the RI error in Eq.~\eqref{eq:loss_func1} yields
\begin{equation}
\mathbf{c}^{\text{RI}} = \mathbf{S}^{-1}\mathbf{w},
\label{eq:invert}
\end{equation}
where $\mathbf{S}$ is the overlap matrix of the periodic basis functions,
\begin{equation}
\begin{aligned}
    S_{i,\sigma}^{j,\tau} & = \sum^{\mathbf{U}_\text{cut}}_\mathbf{U} \sum^{\mathbf{V}_\text{cut}}_\mathbf{V} \braket{\phi_{i,\sigma}(\mathbf{U})}{\phi_{j,\tau}(\mathbf{V})}_\text{u.c.} \\
    & = \sum^{\mathbf{V}_\text{cut}}_\mathbf{V}\braket{\phi_{i,\sigma}(\mathbf{0})}{\phi_{j,\tau}(\mathbf{V})}_{\mathbb{R}^3},
\end{aligned}
\label{eq:S}
\end{equation}
and $\mathbf{w}$ is a vector of the projections of the self-consistent density $\rho^{\text{QM}}(\mathbf{r})$ onto the periodic basis,
\begin{equation}
    w_{i,\sigma} = \sum^{\mathbf{U}_\text{cut}}_\mathbf{U}\braket{\phi_{i,\sigma}(\mathbf{U})}{\rho^{\text{QM}}}_\text{u.c.} = \braket{\phi_{i,\sigma}(\mathbf{0})}{\rho^{\text{QM}}}_{\mathbb{R}^3}.
\label{eq:w}
\end{equation}
Eqs.~\eqref{eq:S} and \eqref{eq:w} display two equivalent expressions for $\mathbf{S}$ and $\mathbf{w}$, which differ in their domain of integration: the subscript u.c. indicates an integration over a single `central' unit cell, $\mathbf{U} = (0,0,0)$, while the the subscript $\mathbb{R}^3$ indicates an integration over all space. We refer to the latter as the `unfolded' representations, which are visualised in Fig.~\ref{fig:method}. (A visualisation of the folded representations can be found in Ref.~\citenum{Knuth2015}). The former expressions describe the `folded' representations, in which contributions to $\mathbf{S}$ and $\mathbf{w}$ from neighboring unit cells are projected onto the `central' unit cell; in practice it is more efficient to evaluate $\mathbf{S}$ and $\mathbf{w}$ in this folded representation. In both representations, the sum over translation vectors can be truncated by defining a cutoff distance from atom $i$ for each basis function $\sigma$, beyond which contributions to $\mathbf{S}$ and $\mathbf{w}$ may be safely neglected; this is indicated by the limits to the sums found in Eqs.~\eqref{eq:S} and \eqref{eq:w}, $\mathbf{U}_\text{cut}$ and $\mathbf{V}_\text{cut}$.

\subsection{Symmetry-adapted learning of three-dimensional electron densities (SALTED)}

The SALTED method predicts the electron density within the RI ansatz defined in Eq.~\eqref{approx_rho}. Rather than using the overlap matrix $\mathbf{S}$ and density projections $\mathbf{w}$ to calculate the RI coefficients of a single structure using Eq.~\eqref{eq:invert}, it uses these quantities obtained from a series of training structures to produce a model which provides approximate expansion coefficients $c_{i,\sigma}$ for related structures, based solely on their nuclear coordinates. In the following, we make use of the RI coefficients themselves as a reference, to disentangle the small error associated with the basis set representation of the electron density described in Eq.~\eqref{approx_rho} from the error that is exclusively associated with the machine-learning approximation.

The SALTED method produces a symmetry-adapted approximation of the expansion coefficients $c_{i,\sigma} \equiv c_{i,n\lambda\mu}$ which mirrors the three-dimensional covariance of the atom-centered spherical harmonics used to expand the density field.~\cite{Grisafi2019} While the property of covariance was first introduced in the space of symmetry-adapted kernel functions,\cite{gris+18prl} one can equally well formulate the problem in the primal space of covariant structural representations. We rely on the general formalism introduced in Ref.~\citenum{will+19jcp}, where an abstract representation of a local environment of the atom $i$ associated with a given structure $A$ is indicated by an abstract ket $\ket{A_i}$ -- thus leaving the freedom to choose any appropriate feature space for the evaluation of the structural representation. In this picture, a generic structural representation of the atomic environment $A_i$ that mirrors the transformation properties of the density-coefficients $c_{i,n\lambda\mu}$ can be constructed by performing the following rotational average~\cite{will+19jcp}:
\begin{equation}\label{eq:covariant}
    \ket{\overline{A_i; \lambda\mu}} = \int d\hat{R}\, \hat{R}\ket{A_i} \otimes \hat{R}\ket{\lambda\mu} ,
\end{equation}
where $\hat{R}$ is a rotation operator, $\ket{\overline{A_i; \lambda\mu}}$ is the symmetry-adapted representation of order $\lambda$, and $\ket{\lambda\mu}$ is an angular momentum state associated with the spherical harmonic $Y_{\lambda}^{\mu}$. At this point, a covariant approximation of the density coefficients could be readily obtained by relying on a linear model based on a set of structural features defined using Eq.~\eqref{eq:covariant}.\cite{gris+19book} However, because the feature-space size can grow rapidly with the complexity of the structural representation, it is typically more convenient to work in the dual space of kernel functions which measure structural similarities between pairs of atomic environments $i$ and $j$ associated with two configurations $A$ and $A'$. From Eq.~\eqref{eq:covariant}, a covariant kernel function can be defined by the braket $\operatorname{k}^\lambda_{\mu\mu'}(A_i,A'_{i'})\equiv\bra{\overline{A'_{i'}; \lambda\mu'}}\ket{\overline{A_i; \lambda\mu}}$. Then, a covariant approximation of the density-coefficients reads as follows:
\begin{equation}\label{eq:coeffs_approx}
    c_{n\lambda\mu}(A_i) \approx \sum_{j\in M}\sum_{|\mu'|\leq \lambda} b_{n\lambda\mu'}(M_j) \operatorname{k}^\lambda_{\mu\mu'}(A_i,M_j)\delta_{a_i a_j} ,
\end{equation}
with $j$ running over a sparse set $M$ of atomic environments that best represent the possible spectrum of structural and chemical variations, while the sum over $\mu'$ expresses the covariant character of the SALTED approximation. $b_{n\lambda\mu'}(M_j)$ are the (covariant) weights we wish to determine upon training the model on a set of $N$ reference densities and atomic configurations. Note that we use the same kernel for all radial channels $n$, and that we have introduced the Kronecker-delta $\delta_{a_i a_j}$ to ensure that only structural environments which are centered about the same atomic species $a$ are coupled. In this work, the actual calculation of the representation $\ket{\overline{A_i; \lambda\mu}}$, or kernel function $\operatorname{k}^\lambda_{\mu\mu'}(A_i,M_j)$, follows the $\lambda$-SOAP formalism first derived in Ref.\citenum{gris+18prl}. However, it is worth pointing out that the construction of Eq.~\eqref{eq:covariant} is in principle general enough to also allow for different functional forms, such as LODE~\cite{gris-ceri19jcp,grisafi2021cs} and NICE.~\cite{niga+20jcp}

Having established a suitable ansatz for approximating the density coefficients, the regression weights $b_{n\lambda\mu}(M_j)$ are determined by the minimization of a loss function which resembles the one used in Eq.~\eqref{eq:loss_func1} to provide a suitable RI approximation of the density field. In particular, given $N$ training configurations and an associated set of reference \textit{ab initio} densities $\{\rho^{\text{QM}}_A(\mathbf{r)}\}$, we can write:
\begin{equation}
\begin{split}
\ell(\bb_{M}) &= \sum_{A=1}^N\int_{u.c.} d\mathbf{r} \left| \tilde{\rho}^\text{ML}_A(\mathbf{r};\bb_{M}) - \rho_A^{\text{QM}}(\mathbf{r}) \right|^2 \\& + \eta\,\bb_{M}^T\mathbf{K}_{{M}{M}}\bb_{M} ,
\label{eq:loss_func2}
\end{split}
\end{equation}
where $\tilde{\rho}^{\text{ML}}(\mathbf{r)}$ is the density approximation that parametrically depends of the regression weights through Eqs.~\eqref{approx_rho} and~\eqref{eq:coeffs_approx}. $\bb_{M}$ indicates a single vector containing the regression weights, whose dimension is determined by the sum of the number of basis functions $(n\lambda\mu)$ in each of the $M$ sparse atomic environments. The kernel matrix $\mathbf{K}_{{M}{M}}$ is defined to be block-diagonal in the atomic types $a$, angular momenta $\lambda$ and radial indexes $n$. Note that a regularization term with an adjustable parameter $\eta$ is introduced in the second line to prevent overfitting the model on the training data.

As detailed in Ref.\citenum{Fabrizio2019}, minimization of Eq.~\eqref{eq:loss_func2} with respect to $b_{n\lambda\mu}(M_j)$ leads to the following regression formula:

\begin{equation}
\bb_{M} = (\mathbf{K}_{{N}{M}}^T\mathbf{S}_{{N}{N}} \mathbf{K}_{{N}{M}} + \eta\mathbf{K}_{{M}{M}})^{-1}\mathbf{K}_{{N}{M}}^T\mathbf{w}_{N} .
\label{eq:weights}
\end{equation}

The vector $\mathbf{w}_{N}$ contains the projections of the training densities on the basis functions (Eq.~\eqref{eq:w}), whose dimension is given by the sum of the number of basis functions $(n\lambda\mu)$ associated with every atomic environment in each of the $N$ training configurations. The matrix $\mathbf{S}_{{N}{N}}$ contains the overlap between the basis functions of each configuration (Eq.~\eqref{eq:S}), and is block-diagonal in the training structures $N$. Note that the overlap matrix is required only to calculate these regression weights; the overlap matrices of target structures are not needed to predict their density-coefficients using Eq.~\eqref{eq:coeffs_approx}. Finally, the rectangular matrix $\mathbf{K}_{{N}{M}}$ contains the kernels which couple the atomic environments of the training structures with those selected to define the sparse approximation of the density coefficients in Eq.~\eqref{eq:coeffs_approx}. Note that the set $M$ is a representative subset of the atomic environments comprising the $N$ training structures, such that we perform a dimensionality reduction commonly known as the \textit{subset of regressors} (SoR) approximation.\cite{quin+05jmlr} In this work, the sparse set is selected using the \textit{farthest point sampling} (FPS) algorithm,\cite{imba+18jcp} using the scalar ($\lambda=0$) SOAP metric.\cite{bart+13prb} 

Using the regression formula of Eq.~\eqref{eq:weights} with the projections of an \textit{all-electron} density directly would result in the major portion of the learning effort being spent on reproducing the core-density peaks at the nuclear positions, especially when considering structures that include heavy atoms. According to Kato's theorem,~\cite{kato1957} however, the form of these peaks in the vicinity of the nuclei is uniquely determined by the nuclear charge, so that one can expect the core-electron contributions to be generally constant across the dataset. To allow the regression to focus solely on the chemically-driven variations of the density field, we provide a baseline value for the vector of density projections $\textbf{w}_{N}$. By averaging just the isotropic ($\lambda=0$) coefficients across the dataset (since all other terms must average to zero for random rotations of the training structures), we obtain a sparse vector of average density coefficients $\bar{\textbf{c}}$ which is used to define the baseline value for the density projections using $\bar{\textbf{w}}=\textbf{S}\bar{\textbf{c}}$. Then, after predicting the variation of the density coefficients $\Delta\textbf{c}_{N}$ relative to this baseline using Eq.~\eqref{eq:coeffs_approx}, the precomputed mean density components $\bar{\textbf{c}}$ are added back to yield the final all-electron density prediction.

From a computational point of view, while the kernel is diagonal across the different types of basis functions $(an\lambda)$, the overlap matrix $\mathbf{S}_{{N}{N}}$ couples these basis functions together, so that the regression in Eq.~\eqref{eq:weights} must be performed on the entire vector of density projections $\mathbf{w}_{N}$. This follows directly from the non-orthogonality of the multi-centered basis set used to expand the electron density as in Eq.~\eqref{approx_rho}. Unlike orthogonal approaches,~\cite{Brockherde2017} this method must deal with regression matrices which quickly become very large with an increasing number of sparse environments $M$ or basis functions. This technical downside is compensated by the great transferability and data-efficiency of the SALTED model, which results from the adoption of a local and symmetry-adapted representation of the scalar field.

\section{Results and Discussion}

\subsection{Validation of the basis}
\label{sec:validation}

We begin by establishing that it is possible to accurately represent the electron density using a linear combination of our chosen basis functions, as described by Eq.~\eqref{approx_rho}. Throughout this paper, we use the so-called ``auxiliary basis functions'' defined in FHI-aims as the basis in which to express the electron density.\cite{Ren2012} This basis set is produced by taking the ``on-site" pair products of the radial parts of the atom-centred numerical orbitals used by FHI-aims in DFT calculations.

A particular set of auxiliary basis functions is therefore defined by the choice of numerical orbitals whose product pairs generate the auxiliary functions. Throughout this work we chose the generating set to be the numerical orbitals used in the calculation of the self-consistent reference density $\rho^{\text{QM}}$, as defined by FHI-aims' ``tight'' settings. However, in general the method presented here does not require this choice of basis function, provided that the basis allows an accurate expansion of the density.\cite{Grisafi2019}

This choice of basis function has two implications for the implementation of the theory laid out in Section \ref{sec:theory}. Firstly, the integrals in Eqs.~\eqref{eq:S} and \eqref{eq:w} are evaluated within FHI-aims using a real-space grid inside an arbitrarily chosen ``central'' unit cell. In order to obtain an efficient implementation when using this ``folded'' representation, care must be taken to find a suitable cutoff for the sum over $\mathbf{U}$ for each individual basis function, since there is significant variation in radial extent between basis functions. Secondly, the near-linear dependencies in the auxiliary basis can result in numerical instability when solving Eq.~\eqref{eq:weights} as the number of basis functions per atom and the number of environments in the sparse set $M$ increase. To ensure a stable solution, this equation is solved with a pseudoinverse calculated using the SVD decomposition, with singular values smaller than $10^{-15}\times\text{dim}(\mathbf{b}_M)$ discarded.

To assess the accuracy of expanding the density in this auxiliary basis, we must calculate the coefficients $\mathbf{c}^{\text{RI}}$ which minimize the error in the approximate density defined in Eq.~\eqref{approx_rho}. These coefficients are given by the RI procedure described in Eqs.~\eqref{eq:loss_func1} and \eqref{eq:invert} and define what we call the RI density, $\tilde{\rho}^\text{RI}$. For a dataset containing $N$ structures, the average percentage error in this density is defined as
\begin{equation}
\bar{\epsilon}^{\text{RI}}_{\rho}(\%) = \frac{100}{N} \sum_A^N \frac{ \int_{u.c.} d\mathbf{r} \left| \tilde{\rho}_A^{\text{RI}}(\mathbf{r}) - \rho^{\text{QM}}_A(\mathbf{r}) \right|} {\int_{u.c.} d\mathbf{r} \rho_A^{\text{QM}}(\mathbf{r})} .
\label{eq:ri_dens_err}
\end{equation}
Note that each term in the sum is normalised by the number of electrons in the structure, making this measure of the error comparable between different systems.

In order to establish the general applicability of the SALTED method, we used three simple test datasets: a metal, a semiconductor and a molecular solid. These datasets consist of:
\begin{enumerate} 
\item An aluminium dataset, containing 50 1-atom unit cells and 50 4-atom unit cells,
\item A silicon dataset, containing 50 2-atom unit cells and 50 16-atom unit cells,
\item The I$_\mathrm{h}$ ice dataset, in which all 100 structures contain four water molecules.
\end{enumerate}
The aluminium structures are generated by randomly varying the positions of the atoms or the lattice vectors of the cell around their equilibrium values. The silicon structures were taken from Ref.~\citenum{Bartok2018}, where they formed part of a dataset used to train a Gaussian approximation potential, and the dataset includes both deformations of the unit cell as well as variations in the atomic positions within the cell. The ice structures were generated using a NPT molecular dynamics trajectory ($P = 1$ atm, $T = 273$ K), with structures sampled every 500 fs. The forces were evaluated using the TIP/4P force field implemented in the LAMMPS software package\cite{Plimpton1995} and the nuclear dynamics were calculated using i-PI.\cite{Kapil2018} Therefore, every dataset includes structures containing significant variations in both the atomic positions and the lattice vectors of the unit cell.

Having defined these datasets, we calculated the error introduced by expressing the density of each structure as a linear combination of auxiliary basis functions. The first column of Table \ref{tab:energy_err_ri} lists the average error in the RI density for each of these datasets, relative to reference densities calculated self-consistently using the local density approximation (LDA).\cite{Hohenberg1964,Kohn1965} In each case, the error is less than 0.1 \%. This compares favourably to previous work, in which Gaussian basis sets were used to express the density of isolated water molecules and simple alkanes and alkenes with mean absolute errors of approximately 0.3 \% and 1 \%, respectively.\cite{Fabrizio2019,Grisafi2019}

\begin{table}
\centering
\setlength{\tabcolsep}{5pt}
\begin{tabular}{c | c c c c c } 
 Dataset & $\bar{\epsilon}_{\rho}^{\text{RI}}$ (\%) & $\bar{\epsilon}^{\text{RI}}_{xc}$ & $\bar{\epsilon}^{'\text{RI}}_{xc}$ & $\bar{\epsilon}^{\text{RI}}_{el}$ & $\bar{\epsilon}^{'\text{RI}}_{el}$ \\
 \hline
 Al & 0.02 & 0.14 & 0.03 & 11.6 & 2.58 \\
 Si & 0.06 & 1.17 & 0.05 & 30.0 & 2.26 \\
 I$_\mathrm{h}$ Ice & 0.01 & 0.00 & 0.00 & 0.19 & 0.01 \\
\end{tabular}
\caption{The average error in the approximate RI density ($\bar{\epsilon}_{\rho}^{\text{RI}}$), along with the average error in exchange-correlation and electrostatic energies derived from it ($\bar{\epsilon}^{\text{RI}}_{xc}$ and $\bar{\epsilon}^{\text{RI}}_{el}$). These errors are relative to the QM reference values. $\bar{\epsilon}^{'\text{RI}}_{xc}$ and $\bar{\epsilon}^{'\text{RI}}_{el}$ are the ``baselined'' average errors, which remain after the mean error has been subtracted from each energy; this indicates the remaining error after the systematic error has been removed. All energies are reported in meV per atom.}
\label{tab:energy_err_ri}
\end{table}

Together with the direct error in $\tilde{\rho}^\text{RI}$, we also investigated the error in the properties derived from this approximate density. By using the Harris energy functional with the LDA, we can compare the exchange-correlation energy derived from an approximate density, $E_{xc}[\tilde{\rho}_A]$, to that derived from a reference density, $E_{xc}[\rho^{\text{ref}}_A]$, for each structure $A$ in a dataset. The absolute error in the exchange-correlation energy per atom is then
\begin{equation}
\epsilon_{xc} = \frac{1}{N_A^{\text{at}}} | E_{xc}[\tilde{\rho}_A] - E_{xc}[\rho^{\text{ref}}_A]|.
\label{eq:energy_err}
\end{equation}
The errors in the exchange-correlation energies derived from the RI density are reported for each dataset in Table \ref{tab:energy_err_ri}, along with the analogous error in the electrostatic energy, $\epsilon_{el}$; the self-consistent density $\rho^{\text{QM}}_A$ is again used as the reference. While charge neutrality is not enforced when obtaining either the RI or the predicted densities, it is very important in the condensed phase to avoid divergent electrostatic terms. In practice, we observe very good charge conservation, with errors typically below $10^{-4}$ $e$ per electron. These small errors are then compensated using the standard practice of applying a constant background charge,\cite{blum+09cpc} which allows us to obtain stable, non-divergent predictions of the electrostatic energy without explicitly normalizing the density predictions.

Table \ref{tab:energy_err_ri} reveals significant variation in the errors of these derived energies between the datasets. The average error introduced to the electrostatic energies of the ice structures is very small ($<$ 1 meV), indicating that the densities produced by expanding in these auxiliary basis functions provide an accurate description of the electrostatic potential. By contrast, the average error introduced to the electrostatic energies of the silicon and aluminium structures are around two orders of magnitude larger, a far greater increase than what one would expect from looking at the error in the density. We find that the corresponding errors in the Hartree energy $E_H$ are much smaller (3.5 meV and 0.2 meV, respectively), indicating that the error in the electrostatic energy arises primarily from the electron-nuclear interaction energy $E_{en}$. As rigorously detailed in the Supporting Information, the error $\delta E_{en}$ associated with the latter contribution is dominated by inaccuracies of the electron density very close to the nuclei, suggesting that an extremely accurate density is required in this region. However, given that the behaviour of the electron density at the atomic positions is expected to be mostly determined by the nuclear charge,~\cite{kato1957} the nature of these errors is largely systematic, as shown in Table \ref{tab:energy_err_ri} by the much smaller ``baselined'' errors which remain after the mean error is subtracted. This suggests that differences in the electrostatic energy are predicted with a far greater accuracy than their absolute values, ensuring the viability of the method for any kind of physical application. Being less sensitive to errors in the electron density localised near the nuclear positions, the average error in the exchange-correlation energies is significantly smaller for each dataset.

Taken together, these observations illustrate an important point: the errors in properties derived from the electron density may depend on the errors in that density in a non-uniform way -- errors in certain regions of space may lead to very large errors in some properties while not significantly increasing the error in other properties. As a result, when approximating the density in this way it may be necessary to find a basis in which to expand the electron density which produces a tolerably low error not only in the density itself, but also some property of interest derived from this density, since the former does not guarantee the latter.~\cite{briling+21arxiv}

\subsection{Predicting electron densities}
\label{sec:Predicting}

Having found a basis in which we can accurately expand the electron density, we then used the SALTED method outlined in Section \ref{sec:theory} to train a model with which to predict the electron densities of our test systems. For this method, the only inputs required are the atomic coordinates, the overlap matrix $\mathbf{S}$ defined in Eqs.~\eqref{eq:S} and electron density projections $\mathbf{w}$ defined in Eq.~\eqref{eq:w} for each structure in the dataset. There are three parameters which must be optimised for each dataset. Two are associated with the $\lambda$-SOAP representations of the configurations in the dataset;\cite{gris+18prl} the other is the regularization parameter $\eta$, which was introduced in the loss function in Eq.~\eqref{eq:loss_func2} to avoid over-fitting. These parameters were optimised using 80 structures in the training set and 20 structures in the validation set. Further details of this optimisation are provided in the Supporting Information.

To assess the accuracy of the electron density predicted by the machine-learning model, we calculated the root mean square difference between the predicted density $\tilde{\rho}^{\text{ML}}$ and the RI density $\tilde{\rho}^{\text{RI}}$, normalised by the standard deviation in the reference densities. The square difference between the predicted and reference densities at point $\mathbf{r}$ for some structure $A$ is given by
\begin{equation}
\left| \tilde{\rho}_A^{\text{ML}}(\mathbf{r}) - \tilde{\rho}_A^{\text{RI}}(\mathbf{r}) \right|^2 = \left| \sum_{i,\sigma,\mathbf{U}} \left( c^{\text{ML}}_{A,i,\sigma} - c^{\text{RI}}_{A,i,\sigma} \right) \phi_{i,\sigma}(\mathbf{r}) \right|^2,
\end{equation}
where the full argument of $\phi_{i,\sigma}$ given in Eq.~\eqref{approx_rho} has been suppressed. Writing $\Delta {c}_{A,i,\sigma} = c^{\text{ML}}_{A,i,\sigma} - c^{\text{RI}}_{i,\sigma}$, we may write the square error in the density of structure $A$ as
\begin{equation}
\begin{aligned}
\left( \epsilon_{\rho,A}^{\text{ML}}\right)^2 & = \int_{u.c.} d\mathbf{r} \left| \tilde{\rho}_A^{\text{ML}}(\mathbf{r}) - \tilde{\rho}_A^{\text{RI}}(\mathbf{r}) \right|^2 \\
& = \Delta \mathbf{c}_A^T \cdot \mathbf{S}_A \Delta \mathbf{c}_A \\
& = \Delta \mathbf{c}_A^T \Delta \mathbf{w}_A,
\end{aligned}
\end{equation}
where $\Delta {w}_{A,i,\sigma} = w^{\text{ML}}_{A,i,\sigma} - w^{\text{RI}}_{A,i,\sigma}$ is the difference between the projections of the predicted and reference density onto the basis function $\sigma$. The standard deviation in the reference density can be written using a similar notation:
\begin{equation}
    s^{\text{RI}}_{\rho} = \sqrt{\frac{1}{N_{t}-1} \sum_A^{N_{t}} \Delta \bar{\mathbf{c}}_A^T \Delta \bar{\mathbf{w}}_A},
\end{equation}
where $\Delta \bar{\mathbf{c}}_A = \mathbf{c}_A^{\text{RI}} - \bar{\mathbf{c}}^{\text{RI}}$ and $\Delta \bar{\mathbf{w}}_A = \mathbf{w}_A^{\text{RI}} - \bar{\mathbf{w}}^{\text{RI}}$, $\bar{\mathbf{c}}^{\text{RI}}$ and $\bar{\mathbf{w}}^{\text{RI}}$ are the mean baseline values for the vectors of coefficients and density projections as defined in Sec.~\ref{sec:theory}, and $N_t$ is the number of structures in the validation set. The percentage root mean square error is then defined as
\begin{equation}
\% \, \text{RMSE} = 100 \times \frac{ \sqrt{ \frac{1}{N_{t}} \sum_A^{N_{t}} \left( \epsilon_{\rho,A}^{\text{ML}}\right)^2}}{s^{\text{RI}}_{\rho} },
\end{equation}
Note that while this definition of the error in the density is similar to the one used in the previous section, it differs in important respects, namely the normalisation factor. In addition, we here use the RI density $\tilde{\rho}^{\text{RI}}$ as the reference density, rather than the self-consistent density $\rho^{\text{QM}}$ as in Section \ref{sec:validation}, since $\tilde{\rho}^{\text{RI}}$ represents the best possible predicted density. This avoids conflating errors arising from the choice of auxiliary basis functions with errors arising from the machine-learning, which could complicate the assessment the accuracy of the machine-learning model.

\begin{figure}[t!]
\centering
\includegraphics[width=8cm]{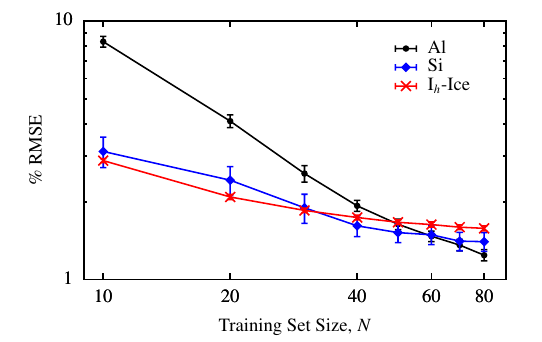}
\caption{Learning curves for each of the test datasets. For each point, the percentage root mean square error is averaged across 10 randomly selected validation sets, each containing 20 structures; the error bars indicate the standard error in the mean.}
\label{fig:learning_curves}
\end{figure}

Having determined the best parameters for each dataset, we calculated learning curves for each of the datasets introduced in Section \ref{sec:validation}. To obtain these learning curves, the error was calculated for 10 randomly selected validation sets each containing 20 structures. The average error across these validation sets is shown in Fig.~\ref{fig:learning_curves} as a function of the training set size. For each dataset, these curves have been converged with respect to the number of reference environments $M$, as demonstrated in the Supporting Information.

For all three datasets, the learning curves indicate a model which reliably and accurately predicts electron densities: in every case the error decreases monotonically as the size of the training set is increased, and for every dataset the error is reduced to below 2\% using just 80 training structures. Note that this is a different definition of the error than that used in in Refs.~\citenum{Fabrizio2019} and ~\citenum{Grisafi2019} when reporting the predicted electron densities of isolated molecules and dimers; using the same metric as in those works, we find errors of at most 0.15\%, lower than those obtained for isolated molecules. It is clear that there is no significant loss of accuracy from extending the SALTED method introduced in Ref.~\citenum{Grisafi2019} to periodic systems and using a numeric atom-centered basis set representation.

\begin{figure}
\includegraphics[width=8cm]{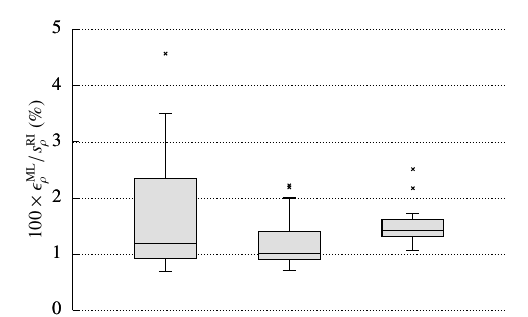} \\
\includegraphics[width=8cm]{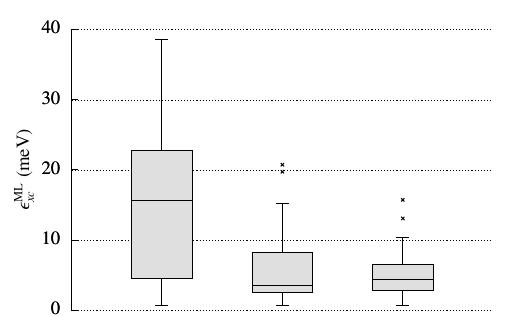} \\
\includegraphics[width=8cm]{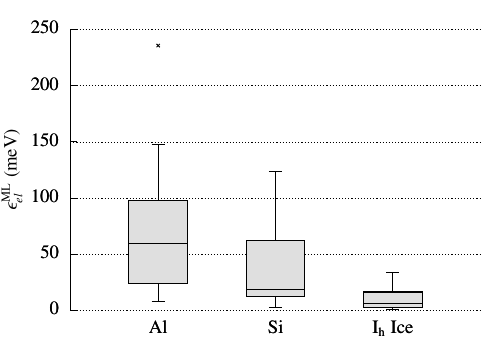}
\caption{The upper panel shows the distribution of the percentage root square errors in the density, $100 \times \epsilon^{\text{ML}}_{\rho} / s^{\text{\text{RI}}}_{\rho}$, arising from the predicted density of 20 randomly selected structures from each dataset. The lower two panels show the distribution of the absolute errors in the exchange-correlation energy, $\epsilon^{\text{ML}}_{xc}$, and electrostatic energy, $\epsilon^{\text{ML}}_{el}$, for the same structures.}
\label{fig:RMSE_xc_el}
\end{figure}

We again used the Harris functional with the LDA to calculate the exchange-correlation and electrostatic energies associated with each predicted density, $\tilde{\rho}_A^{\text{ML}}$. The errors in the energies are once more given by Eq.~\eqref{eq:energy_err}, using $\tilde{\rho}_A^{\text{RI}}$ as the reference density as we did when evaluating the errors in the predicted density directly. The distributions of these errors are shown in Figure \ref{fig:RMSE_xc_el}, along with the distribution of the percentage errors in the density $100 \times \epsilon^{\text{ML}}_{\rho} / s^\text{RI}_{\rho}$ for the particular set of test structures used to calculate the energies.

Firstly, it is clear that while the average error in the density is similar for all three datasets, there is a larger variation in the error in the density between structures in the aluminium dataset. These outliers are primarily due to the 1-atom unit cells, contained only in the aluminium dataset, in which the effect of a single poorly described environment on the error is magnified relative to systems containing more atoms. By contrast, the predicted densities of the silicon and ice structures are consistently accurate, with an error of below 2\% for almost all of the structures.

As might be expected, this trend is reflected in the exchange-correlation energies derived from the electron densities, with a much wider distribution of errors for the aluminium dataset than the silicon and ice datasets. By contrast, the distribution of errors in the electrostatic energies is not much broader for aluminium than for silicon, although the median error is significantly larger. One possible reason for this behaviour will be discussed in Section \ref{sec:Extrap}. For the ice dataset, the median errors in both the electrostatic and exchange-correlation energies are approximately 5 meV per atom, as is the median error in the exchange-correlation energy for the silicon dataset. These errors indicate that, in general, the densities produced by the SALTED method are sufficiently accurate to provide reasonable estimates of energies derived from those densities from just a small number of training structures, with no information about those properties built into the training model. We will discuss some possible routes to further improving the accuracy of these derived quantities in the conclusions.

\begin{table}
\centering
\setlength{\tabcolsep}{5pt}
\begin{tabular}{c | c c c c } 
 Dataset & $\bar{\epsilon}^{\text{ML}}_{xc}$(I) &  $\bar{\epsilon}^{\text{ML}}_{xc}$(D) & $\bar{\epsilon}^{\text{ML}}_{el}$(I) &  $\bar{\epsilon}^{\text{ML}}_{el}$(D) \\
 \hline
 Al & 15.4 & 2.85 & 68.2 & 25.4 \\
 Si & 6.30 & 2.20 & 37.0 & 108 \\
 I$_\mathrm{h}$ Ice & 5.41 & 2.05 & 10.0 & 6.38 \\
\end{tabular}
\caption{The mean absolute errors in the exchange-correlation and electrostatic energies ($\bar{\epsilon}^{\text{ML}}_{xc}$ and $\bar{\epsilon}^{\text{ML}}_{el}$) derived from the predicted electron densities (the indirect errors, I), compared to the mean absolute errors observed when those energies are predicted directly using Gaussian process regression (D) for each of the three datasets. These errors are relative to the RI references values. All energies are reported in meV per atom.}
\label{tab:indirect_vs_direct}
\end{table}

Finally, we compare the accuracy of the energies calculated in this ``indirect'' manner to the ones predicted ``directly'' using a simple Gaussian process regression model, again using 80 training structures. These GPR models are optimised independently of the SALTED models used to predict the electron densities. The mean absolute errors in the electrostatic and exchange-correlation energies predicted by both the indirect (I) and direct (D) methods are shown in Table \ref{tab:indirect_vs_direct}. The errors arising from the indirect predictions are larger than those arising from the direct predictions, but all are of comparable magnitude. (The direct GPR model used to predict the electrostatic energies of silicon appears to suffer from numerical instability arising from the small number of training points, resulting in the anomalous result in Table \ref{tab:indirect_vs_direct}.) Furthermore, a separate GPR model must be optimised for each property of interest in order to learn them directly (the resulting hyper-parameters are provided in the Supporting Information). By contrast, both energies are obtained from a single model when calculated indirectly, along with any other electronic property of interest which may be derived from the predicted electron density. By predicting the electron density, the SALTED method effectively allows the prediction of a wide range of properties simultaneously.

\subsection{Extrapolating electron densities}
\label{sec:Extrap}
In the previous section, we demonstrated that the electron density of periodic systems could be accurately predicted using training data generated from similar structures. However, the real utility in local machine-learning algorithms such as the one presented here is the ability to accurately and efficiently predict properties of systems which are challenging and expensive to obtain using a direct \emph{ab initio} calculation. Therefore, we would like to establish whether a machine-learning model trained on smaller systems is able to accurately predict the electron densities of larger systems containing similar chemical environments.

To investigate this, we turned to a more realistic example: predicting the electron densities of I$_\mathrm{h}$ ice supercells using the SALTED model trained on the 4-molecule cells described in Section \ref{sec:validation}. To generate a representative test set, we ran MD simulations of ice supercells containing 64, 128, 256 and 512 molecules (up to 1536 atoms) under the same conditions used to generate configurations for the training dataset, and sampled the resulting trajectories every 200 fs following a 5 ps equilibration to obtain twenty independent configurations at each cell size. We then predicted the electron densities of each of these structures using a SALTED model trained using all 100 structures of the 4-molecule ice dataset, and calculated the exchange-correlation and electrostatic energies derived from these extrapolated densities.

\begin{figure}
    \centering
    \includegraphics[width=8cm]{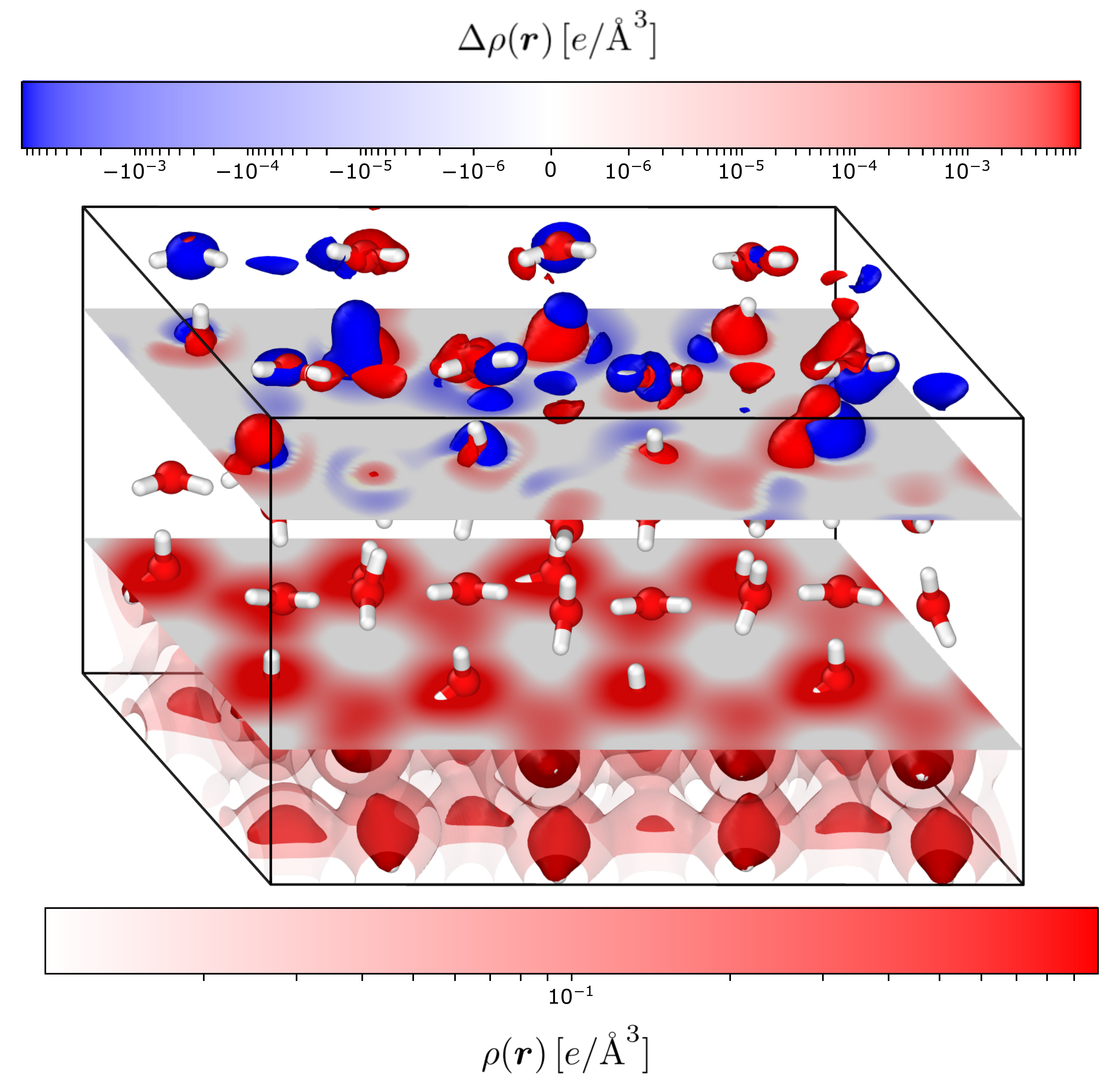}
    \caption{A two-dimensional cut of the predicted electron density of a 64-molecule ice supercell (\textit{lower slice and colorbar}) and of the error in the density with respect to the reference DFT calculation (\textit{upper slice and colorbar}). The figure also reports the corresponding three-dimensional contours at isovalues of 1.0, 0.1, 0.01$e$/{\AA$^3$} and $\pm$0.001$e$/{\AA$^3$} respectively.}
    \label{fig:ice_dens}
\end{figure}

Fig.~\ref{fig:ice_dens} contrasts the error in the predicted density $\tilde{\rho}^\text{ML}$ with the self-consistent electron density $\rho^\text{QM}$ in slices in the $xy$-plane of a 64-molecule supercell. This illustrates that the errors introduced by the SALTED method are an extremely small fraction of the total electron density - note the different scales on the two colourbars. In addition, there is no clear pattern in the errors in the predicted density, suggesting that the training data obtained from just 100 small structures contains sufficient information to avoid introducing large systematic errors into the extrapolated density.

\begin{figure}
\centering
\includegraphics[width=8cm]{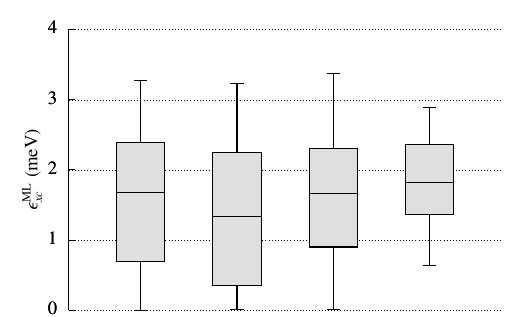} \\
\includegraphics[width=8cm]{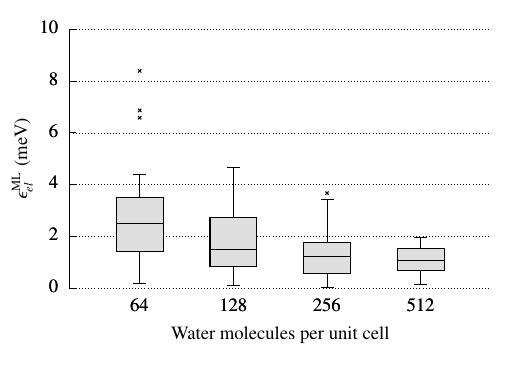} 
\caption{The distribution of the absolute errors in the exchange-correlation energy, $\epsilon^{\text{ML}}_{xc}$, and electrostatic energy, $\epsilon^{\text{ML}}_{el}$, arising from the predicted density of 20 ice supercells containing increasing numbers of water molecules.}
\label{fig:ice_extrap}
\end{figure}

We measure the accuracy of the energies derived from the extrapolated density relative to those derived from the self-consistent density, $\rho^{\text{QM}}$. While in principle this convolves the errors introduced by the SALTED method with the errors introduced by the choice of basis functions, the latter have been shown to be extremely small (Table \ref{tab:energy_err_ri}). Therefore, the errors presented here are dominated by errors introduced by the SALTED method, and provide a reasonable measure of the accuracy of this method. The errors in the energies derived from the extrapolated electron densities are shown in Fig.~\ref{fig:ice_extrap}.

\begin{table}
\centering
\begin{tabular}{c | c c c c } 
 Molecules & $\bar{\epsilon}^{\text{ML}}_{xc}$(I) &  $\bar{\epsilon}^{\text{ML}}_{xc}$(D) & $\bar{\epsilon}^{\text{ML}}_{el}$(I) &  $\bar{\epsilon}^{\text{ML}}_{el}$(D) \\
 \hline
 64 & 1.57 & 2.25 & 2.90 & 8.19 \\
 128 & 1.29 & 3.21 & 1.80 & 8.82 \\
 256 & 1.66 & 3.67 & 1.41 & 9.63 \\
 512 & 1.82 & 3.60 & 1.09 & 9.51 \\
\end{tabular}
\caption{The mean absolute errors in the exchange-correlation and electrostatic energies ($\bar{\epsilon}^{\text{ML}}_{xc}$ and $\bar{\epsilon}^{\text{ML}}_{el}$) derived from the predicted electron densities (the indirect errors, I), compared to the mean absolute errors observed when those energies are predicted directly using Gaussian process regression (D) for each size of ice supercell. These errors are relative to the QM reference values. All energies are reported in meV per atom.}
\label{tab:indirect_vs_direct_extrap}
\end{table}

\begin{figure}[t!]
\centering
\includegraphics[width=8cm]{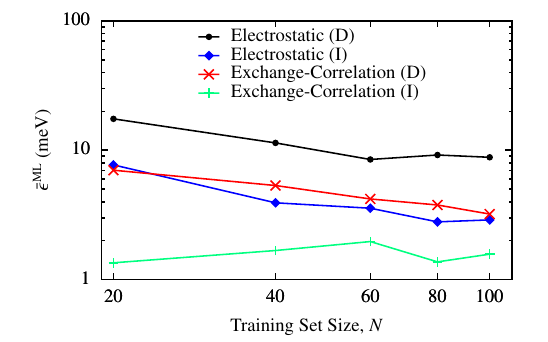}
\caption{Learning curves for the mean absolute errors in the exchange-correlation and electrostatic energies ($\bar{\epsilon}^{\text{ML}}_{xc}$ and $\bar{\epsilon}^{\text{ML}}_{el}$) derived from the predicted electron densities (the indirect errors, I), and predicted directly using Gaussian process regression (D) for the 64-molecule ice supercells. These errors are relative to the QM reference values. Equivalent plots for the other supercell sizes can be found in the Supporting Information.}
\label{fig:direct_learning_curves}
\end{figure}

It is clear that the quality of the electron densities predicted for large ice supercells does not introduce an increase of the error on the derived energies with increasing system size. The predicted exchange-correlation and electrostatic energies are within 5 meV of the converged energy for almost every structure at every system size. This clearly demonstrates the power of our local machine-learning approach: the ground state electron density of large systems can be accurately predicted using information straightforwardly obtained from a small number of structures each containing 100 times fewer atoms than the target system. In fact, Table \ref{tab:indirect_vs_direct_extrap} demonstrates that the energies predicted in this indirect manner are more accurate than those obtained using the GPR model optimised on the 4-molecule ice structures to directly predict the energies. Fig.~\ref{fig:direct_learning_curves} shows the learning curves for the energies of the 64-molecule supercells, as predicted using both the direct and indirect machine learning methods. Interestingly, the indirect predictions of the electrostatic and exchange-correlation energy contributions do not necessarily show a monotonic decrease in the error as a function of the training set size. This is reminiscent of what was observed for SA-GPR predictions of molecular dipole moments when extrapolating to larger compounds than those trained against,\cite{veit+20jcp} suggesting that models trained on small systems can develop weights corresponding to long-range correlations that are damaging to the extrapolative prediction. Furthermore, since SALTED predicts the electron density directly, rather than these energies, it is not clear that one should expect the derived energies to exhibit perfectly monotonically decreasing learning curves. Nevertheless, for every number of training points, the indirect method shows superior performance; the same qualitative behaviour is observed for every supercell size, with the remaining learning curves included in the Supporting Information.

Moreover, this indicates that the information contained within our local learning model is sufficient to describe the relevant local atomic environments, regardless of the number of atoms in the system; any error introduced by finite size effects appears to be smaller than the error introduced by the model itself. In fact, if anything, the per atom error in the electrostatic energy appears to decrease with increasing system size. This may be the result of a cancellation between contributions to the error in the electrostatic energy of opposite signs; as the system size increases, the probability that these contributions to the error cancel one another out increases, lowering the error in the electrostatic energy per atom. This also helps to explain the distribution of the absolute errors in the electrostatic energy for the validation datasets shown in Fig.~\ref{fig:RMSE_xc_el} - the aluminium dataset contains the structures with fewest atoms, followed by the silicon dataset, while all of the ice structures contain 12 atoms, and the median error decreases in the same order.

\subsection{Learning heterogeneous datasets}

In the previous sections we have established the accuracy of the SALTED method for predicting the electron densities and derived properties of chemically homogeneous datasets. We now put our approach to the test for heterogeneous datasets. To investigate the additional challenges these introduce, we analysed two further scenarios. The first is simply the amalgamation of the three homogeneous datasets introduced in Section \ref{sec:validation}; this will be refered to as the ``mixed'' dataset. The second consists of hybrid organic-inorganic perovskites (HOIP) selected from the dataset published in Ref.~\citenum{Kim2017}. These structures have a common motif of three F atoms and one Sn atom, along with small organic molecules which varied between the different configurations. These small molecules are composed of some combination of C, N and H atoms. This dataset presents a far greater challenge than any of those previously considered: each structure contains at least 4 different atomic species, one of which is a heavy transition metal. Furthermore, this dataset contains a total of just 100 structures, which allows a direct comparison with the homogeneous datasets above.

Having defined the HOIP dataset, we followed the same procedure outlined in Sections \ref{sec:validation} and \ref{sec:Predicting} to validate our choice of auxiliary basis for these systems. After calculating the RI density $\tilde{\rho}^\text{RI}$ for each structure, we find an average error in the density of $\bar{\epsilon}^{\text{RI}}_{\rho} = 0.3\%$, following the definition in Eq.~\eqref{eq:ri_dens_err}. This is a little larger than the errors observed for the homogeneous datasets in Table \ref{tab:energy_err_ri}, but is still comparable to the errors observed in previous literature.\cite{Fabrizio2019,Grisafi2019} This small increase in the average error in the RI density does not lead to an increase in the corresponding average errors in the electrostatic and exchange-correlation energies derived from this density, 11.5 meV and 0.07 meV respectively. In addition, this basis allows excellent charge conservation, introducing errors of up to just $10^{-7}$ $e$ per electron. Therefore, we are satisfied that the numerical auxiliary basis used in FHI-aims provides an accurate expansion of the densities of these perovskite structures.

\begin{figure}[t!]
\centering
\includegraphics[width=8cm]{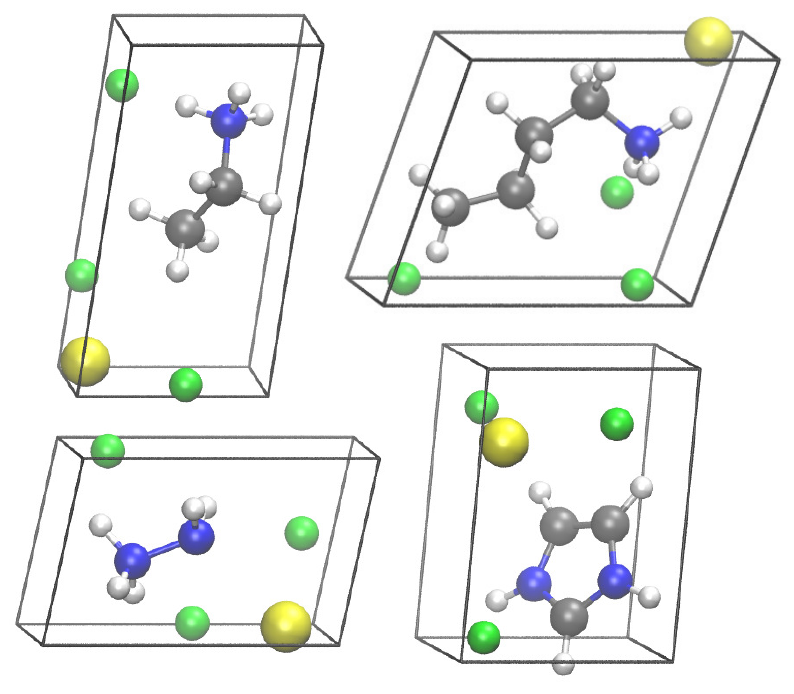} \\
\includegraphics[width=8cm]{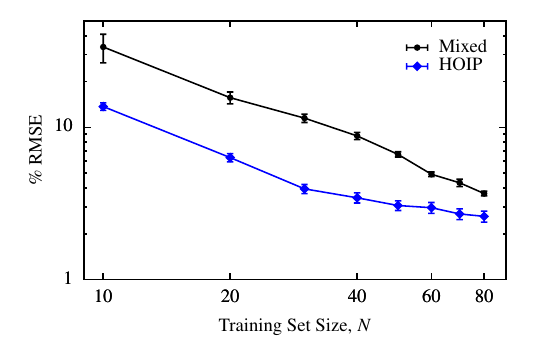} \\
\caption{Above: A selection of the hybrid organic-inorganic perovskite structures. Tin atoms are shown in yellow, fluorine atoms in green, nitrogen atoms in blue, carbon atoms in grey and hydrogen atoms in white. Below: The learning curves for the mixed and HOIP datasets. For each point, the percentage root mean square error is averaged across 10 randomly selected validation sets, each containing 20 structures; the error bars indicate the standard error in the mean.}
\label{fig:hetero_learning_curve}
\end{figure}

We then built SALTED models with which to predict the electron densities of structures within these two heterogeneous datasets. The hyperparameters for both of these models were optimised as outlined in Section \ref{sec:Predicting}, with further details and the selected hyperparameters provided in the Supporting Information. The resulting learning curves are shown in Fig.~\ref{fig:hetero_learning_curve}, and are analogous to those shown in Fig.~\ref{fig:learning_curves}. For both datasets we observe remarkably accurate results with just 80 training structures, with errors below 4\% for the mixed dataset and below 2\% for the HOIP dataset. It is also clear that in both cases the learning curves have not saturated, as might be expected when using so few training structures to describe these heterogeneous dataset. Therefore, applying SALTED to more complex datasets will, with sufficient training structures, produce models that present an accuracy comparable to that achieved for the chemically-homogeneous datasets. However, ramping up the number of training structures $N$ would require a larger number of sparse environments $M$ to represent a richer spectrum of chemical variations, which in the present formalism would imply reaching a computational bottleneck given by the requirement to store and invert prohibitively large matrices. In particular, when using the tight basis sets of FHI-AIMS, working with a number of sparse environments $M\sim10^3$ would mean inverting matrices with dimensions larger than $10^5\times10^5$. A possible solution to this problem would be to avoid finding the explicit solution of the regression problem, instead directly minimizing the loss function of Eq.~\eqref{eq:loss_func2}. This and other appropriate strategies to circumvent the unfavorable scaling of the training procedure with the system size will be the subject of future investigation.

The charge conservation of the electron densities of Al and Si predicted using the SALTED model trained on the mixed database is of the same order as those predicted by the models trained on the separate datasets, with the RMSE charge conservation error rising from $3\times 10^{-5}$ to $6\times 10^{-5}$ $e$ per electron for Al, and from $9\times 10^{-6}$ to $2\times 10^{-5}$ $e$ per electron for Si. By contrast, there is a notable increase in the charge integration error for the electron densities of ice predicted by this model, with the RMSE rising from $1\times 10^{-4}$ to $8\times 10^{-4}$ $e$ per electron using the SALTED model trained on the mixed dataset, and even larger errors of around $8\times 10^{-3}$ $e$ per electron are seen for the HOIP dataset. This can be understood by considering the role of the SOAP hyper-parameters. For the individual homogeneous datasets, different optimal hyperparameters are obtained for each set. However, when considering the mixed dataset, a single set of hyperparameters must be chosen to describe the whole set. We find that the optimal parameters for the mixed set are closer to the optimal parameters for Al and Si than for ice, which then results in a significant deterioration of the charge conservation for the latter structures. This problem is even more pronounced for the HOIP dataset. In fact, the optimization of the ML hyper-paramaters is largely dominated by the presence of the Sn atoms, leading to a very smooth definition of the SOAP atom-density that is used as a structural descriptor, i.e., $r_\text{c}=13$~{\AA} and $\sigma=0.9$~{\AA}. A strategy to solve this issue would consist of using different spatial resolutions for the SOAP description depending on the more or less diffuse nature of the density components to be predicted.

The problem of charge conservation results in significant errors in the energies derived from the predicted densities. While the average error in the indirect predictions of the electrostatic and exchange-correlation energies of Al and Si increases only by a factor of $\sim$2 relative to the indirect predictions of the separate models, this rises to a factor of $\sim$5 for ice. Not surprisingly, these errors are larger in the indirectly predicted energies of the HOIP structures - on the order of 1 eV. Increasing the training set size and adopting  a ML description that can be fine-tuned to represent different kind of density components will therefore be essential to obtain accurate predictions in similar highly heterogenous datasets. 

\section{Conclusions}

We have shown how to use SALTED models to accurately predict the electron density of condensed-phase systems. The locality of the machine-learning model is reflected in the local nature of the atom-centred expansion of the density field, which is made possible for periodic systems through the use of numerical auxiliary basis as implemented in FHI-aims. The adopted RI basis comes along with an accurate decomposition of the density, yielding negligible basis set errors, and is tunable in response to the accuracy required for a particular system. While the non-orthogonality of these basis functions results in regression models which rapidly increase in dimensionality with the size of the basis, this cost is offset by the transferability and locality of the resulting models. As already proven in other contexts,~\cite{Grisafi2019,Fabrizio2019,briling+21arxiv} the local nature of the approach allows for accurate linear-scaling predictions, enabling a massive increase in the size of the systems under study.

We tested the method by training three models which predicted the electron density of a metal, a semiconductor and a molecular solid, finding that in all cases a stable, reliable and accurate model for the density was produced, with a RMSE below 2\% obtained for each validation set using fewer than 100 training structures. We established that exchange-correlation energies could be derived from these densities, with errors lying below 10 meV per atom for the majority of structures. We then demonstrated the ability of SALTED to predict the electron density of very large systems employing a model trained on much smaller systems, using training data obtained from ice cells containing just 4 molecules to predict the density of cells containing 64, 128, 256 and 512 molecules. We found no loss of accuracy as the system size increased, illustrating the ability of the SALTED method to obtain accurate electron densities of large crystalline systems without the need for a self-consistent DFT calculation. In particular, the errors in the derived electronic energies do not increase with increasing system size. Furthermore, these derived energies are more accurate than those predicted using a direct machine-learning model also trained only on the small unit cells. Finally, we used SALTED to predict the electron densities of two heterogeneous datasets, and found predicted densities with RMSE below 4\%. However, in order to derive reliable energies from these heterogeneous datasets, technical developments are required to firstly store larger matrices related to the larger number of sparse environments $M$ required for these scenarios, and secondly to fine tune the SOAP hyper-parameters to account for the different atomic sizes found in heterogeneous datasets.

As recently demonstrated in Ref.~\citenum{briling+21arxiv}, the metric chosen to define the the RI and machine-learning approximation can impact the quality of density-derived properties. Formulations discussed in Ref.~\citenum{briling+21arxiv} could be pursued in order to obtain more accurate derived electrostatics energies. Understanding whether or not this choice would compromise the quality of the exchange-correlation energy is matter of future investigation. Another possible improvement for all-electron densities would be to generate data to train the model that is based only on density changes with respect to, for example, a superposition of free atom densities. This procedure could mitigate any problems representing the density close to the nuclei, which led to the errors in the electrostatic energy discussed in Section \ref{sec:validation}.

An extension of the method will be needed to treat those systems that are dominated by non-local physical effects. In fact, while the locality of SOAP-based and similar representations is crucial for allowing transferable predictions in very heterogeneous datasets,~\cite{wilk+19pnas} the accurate description of highly polarizable surfaces and/or ionic systems necessarily requires the spatial nearsightedness of the learning model to be overcome. In this regard, integrating SALTED with long-range representations of the atomic structure~\cite{grisafi2021cs} that can be properly combined with short-range, many-body features~\cite{niga+20jcp} will represent an attractive possibility for enabling the accurate prediction of electron densities in response to far-field perturbations. In addition, work is underway to incorporate the prediction of density gradients along with electron densities, allowing the indirect prediction of electronic properties through generalized-gradient approximation functionals.

In perspective, the application of SALTED to periodic systems paves the way for inexpensive prediction of the electron densities of bulk liquids and solids which can be directly probed by experimental techniques, e.g., X-ray scattering experiments.~\cite{korits+01cr} The possibility of treating on an equal footing both molecular, crystalline and metallic systems represents a great advantage in the computational study of heterogeneous materials, such as those involved in catalytic reactions and electrochemical processes.

\section{Supporting Information, source code and data availability}
The Supporting Information contains the equations for the calculation of the overlap matrix and vector of projections in periodic systems, an error analysis of the electrostatic and Hartree energies, the optimisations of the SALTED hyper-paramaters for the homogenous and heterogenous datasets, the optimisation of the direct GPR hyper-parameters and their learning curves, and an illustration of the application of  SALTED to isolated molecules using NAOs.

A Python based implementation of the SALTED method is free to download at \url{https://github.com/andreagrisafi/SALTED/tree/legacy}. The implementation to obtain and read the quantities necessary for the GPR presented here is available as part of the FHI-aims code package \url{https://fhi-aims.org}, and all the data required to reproduce the results is available at \url{https://dx.doi.org/10.17172/NOMAD/2021.06.07-1}.

\section{Acknowledgments}

This work was supported by the Max Planck-EPFL Center for Molecular Nanoscience and Technology. A.M.L. is supported by the Alexander von Humboldt Foundation. M.R. acknowledges funding through Deutsche Forschungsgemeinschaft Projektnummer 182087777-SFB 951.

\bibliography{refs}

\end{document}


\tableofcontents

\section{Calculation of the overlap matrix and vector of projections in periodic systems}

In Eqs.~4 and 5 of the main text we define the overlap matrix of the basis functions, $\mathbf{S}$, and the vector containing the projection of the density onto each basis function, $\mathbf{w}$, in compact notation. We gave two equivalent forms of these objects - the ``folded'' representation, in which the domain of integration is restricted to a single unit cell onto which contributions from neighbouring unit cells are projected, and the ``unfolded'' representation in which the integral is performed over all space. Here we present those same expressions with all arguments made explicit:
\begin{equation}
\begin{aligned}
S_{i,\sigma}^{j,\tau} & = \int_{u.c.} d\mathbf{r} \sum_{\mathbf{U,V}} \phi_{i,\sigma}(\mathbf{r} - \mathbf{R}_i + \mathbf{T}(\mathbf{U})) \phi_{j,\tau}(\mathbf{r} - \mathbf{R}_j + \mathbf{T}(\mathbf{V})) \\
& = \int_{\mathcal{R}^3} d\mathbf{r} \sum_{\mathbf{U}} \phi_{i,\sigma}(\mathbf{r} - \mathbf{R}_i + \mathbf{T}(\mathbf{U})) \phi_{j,\tau}(\mathbf{r} - \mathbf{R}_j)
\end{aligned}
\end{equation}
\begin{equation}
\begin{aligned}
w_{i,\sigma} & = \int_{u.c.} d\mathbf{r} \sum_{\mathbf U} \phi_{i,\sigma}(\mathbf{r} - \mathbf{R}_i + \mathbf{T}(\mathbf{U})) \rho^{\text{QM}}(\mathbf{r}), \\
& = \int_{\mathcal{R}^3} d\mathbf{r} \, \phi_{i,\sigma}(\mathbf{r} - \mathbf{R}_i) \rho^{\text{QM}}(\mathbf{r}),
\end{aligned}
\end{equation}
The equivalence of the two expressions for $w_{i,\sigma}$ can be concisely proven, beginning with the unfolded representation and dividing the integral over all space into an sum of integrals over every unit cell:
\begin{equation}\begin{split}
    w_{i,\sigma} & = \int_{\mathcal{R}^3} d\mathbf{r} \, \phi_{i,\sigma}(\mathbf{r} - \mathbf{R}_i + \mathbf{T}(\mathbf{0})) \rho^{\text{QM}}(\mathbf{r}) \\
    & = \int_{\mathbf{U}=(0,0,0)} d\mathbf{r} \phi_{i,\sigma}(\mathbf{r} - \mathbf{R}_i+ \mathbf{T}(\mathbf{0})) \rho^{\text{QM}}(\mathbf{r})  + \int_{\mathbf{U}=(1,0,0)} d\mathbf{r} \phi_{i,\sigma}(\mathbf{r} - \mathbf{R}_i + \mathbf{T}(\mathbf{0})) \rho^{\text{QM}}(\mathbf{r})  + \dots \\
    & = \sum_\mathbf{U} \int_\mathbf{U} d\mathbf{r} \, \phi_{i,\sigma}(\mathbf{r} - \mathbf{R}_i + \mathbf{T}(\mathbf{0})) \rho^{\text{QM}}(\mathbf{r}) \\
    & = \int_{\mathbf{U}=(0,0,0)} d\mathbf{r} \sum_{\mathbf U} \phi_{i,\sigma}(\mathbf{r} - \mathbf{R}_i + \mathbf{T}(\mathbf{U})) \rho^{\text{QM}}(\mathbf{r}),
\end{split}
\end{equation}
Here the subscript to the integral $\mathbf{U}$ indicates that the domain of integration is the unit cell translated by an integer multiple $\mathbf{U} = (U_x,\, U_y,\, U_z)$ of the lattice vectors from the central reference unit cell. The equivalence of the two representations of the overlap matrix can be proven in an analogous manner.

\section{Error analysis of the electrostatic and Hartree energies}

The electrostatic energy is defined as
\begin{equation}
    E_{el} = \frac{1}{2} \int d\mathbf{r} \int d\mathbf{r}' \frac{\rho(\mathbf{r})\rho(\mathbf{r}')}{\left| \mathbf{r} - \mathbf{r}' \right|} - \int d\mathbf{r} \rho(\mathbf{r}) \sum_i \frac{Z_i}{ \left|\mathbf{r} - \mathbf{R}_i \right|}
\label{eq:el}
\end{equation}
where the sum over $i$ runs over all atoms in the system with $\mathbf{R}_i$ the position of atom $i$. The Hartree energy $E_H$ is defined as the first term of Eq.~\eqref{eq:el}, the electron-electron contribution to the electrostatic energy. To first order, errors in the electron density $\delta \rho(\mathbf{r})$ introduce the following errors to the electrostatic and Hartee energies:
\begin{gather}
    \delta E_{el} = \int d\mathbf{r} \int d\mathbf{r}' \frac{\delta \rho(\mathbf{r}) \rho(\mathbf{r}')}{\left| \mathbf{r} - \mathbf{r}' \right|} - \int d\mathbf{r} \delta \rho(\mathbf{r}) \sum_i \frac{Z_i}{ \left|\mathbf{r} - \mathbf{R}_i \right|} \\
    \delta E_{H} = \int d\mathbf{r} \int d\mathbf{r}' \frac{\delta \rho(\mathbf{r}) \rho(\mathbf{r}')}{\left| \mathbf{r} - \mathbf{r}' \right|}
\end{gather}
The error in the electrostatic energy can be re-written to simplify a direct comparison between the two errors:
\begin{equation}
\begin{aligned}
    \delta E_{el} & = \int d\mathbf{r} \delta \rho(\mathbf{r}) \left[\int d\mathbf{r}' \frac{ \rho(\mathbf{r}')} {\left| \mathbf{r} - \mathbf{r}' \right|} -  \sum_i \frac{Z_i}{ \left|\mathbf{r} - \mathbf{R}_i \right|} \right] \\
    & = \int d\mathbf{r} \delta \rho(\mathbf{r}) \left[\int d\mathbf{r}' \frac{ \rho(\mathbf{r}')}{\left| \mathbf{r} - \mathbf{r}' \right|} -  \sum_i Z_i \frac{\delta(\mathbf{r} - \mathbf{r}')}{ \left|\mathbf{r} - \mathbf{R}_i \right|} \right] \\
    & = \int d\mathbf{r} \delta \rho(\mathbf{r}) \left[\int d\mathbf{r}' \frac{ \rho(\mathbf{r}')}{\left| \mathbf{r} - \mathbf{r}' \right|} -  \sum_i Z_i \frac{\delta(\mathbf{r} - \mathbf{r}')}{ \left|\mathbf{r} - \mathbf{R}_i \right|} \right] \\
    & = \int d\mathbf{r} \delta \rho(\mathbf{r}) V(\mathbf{r})
\end{aligned}
\end{equation}
where $V(\mathbf{r})$ is the electrostatic potential. This demonstrates that provided the contributions to the error from the electron-electron interactions and electron-nuclear attractions are of the same order, these two terms will screen one another, reducing the error in the electrostatic energy relative to the error in the Hartree error. This is in fact what we observe in the errors arising from the predicted densities described in Section IIIB of the main text. These are summarised in the last two columns Table.~\ref{tab:predict_errs}.

\begin{table}[t]
\centering
\begin{tabular}{c | c c c c } 
 Dataset & $\bar{\epsilon}^\text{RI}_{el}$  & $\bar{\epsilon}^\text{RI}_H$ & $\bar{\epsilon}^\text{ML}_{el}$ & $\bar{\epsilon}^\text{ML}_H$ \\
 \hline
 Al & 11.6 & 0.2 & 68.2 & 230.0 \\
 Si & 30.0 & 3.5 & 37.0 & 56.7 \\
 I$_\mathrm{h}$ Ice & 0.19 & 0.01 & 10.0 & 24.2 \\
\end{tabular}
\caption{The average errors in the electrostatic and Hartree energies derived from the RI density and the densities predicted using the SALTED method (ML) for the three validation. All errors are in meV.}
\label{tab:predict_errs}
\end{table}


However, we observe the opposite trend in the errors arising from the RI densities described in Section IIIA - the errors in the electrostatic energy are significantly larger than those in the Hartree energy, as shown in the first two columns of Table.~\ref{tab:predict_errs}. This requires that:
\begin{equation}
    \int d\mathbf{r} \, \delta \rho(\mathbf{r}) \sum_i \frac{Z_i} { \left|\mathbf{r} - \mathbf{R}_i \right|} \gg \int d\mathbf{r} \, \delta \rho(\mathbf{r}) \int d\mathbf{r}' \frac{ \rho(\mathbf{r}')}{\left| \mathbf{r} - \mathbf{r}' \right|}.
\label{eq:condition}
\end{equation}
We can show that this condition is satisfied when the error in the density is dominated by errors near the nucleus, such that to a first approximation $\delta\rho(\mathbf{r}) = \varepsilon \sum_j \delta(\mathbf{r} - \mathbf{R}_j)$. Inserting this into Eq.~\eqref{eq:condition} yields
\begin{equation}
   \varepsilon \sum_{ij}  \frac{Z_i}{ \left|\mathbf{R}_j - \mathbf{R}_i \right|} \gg  \varepsilon \int d\mathbf{r}' \, \sum_j \frac{\rho(\mathbf{r}')}{\left| \mathbf{R}_j - \mathbf{r}' \right|}.
\end{equation}
The left hand side will be dominated by terms where $i=j$, while the contributions to the integral on the right hand side will be largest when $\mathbf{r}' = \mathbf{R}_j$. For these contributions, the only difference between the two terms are the numerators: the nuclear charge of nucleus $j$ $Z_j$ on the left compared with the electron density at nucleus $j$ on the right. The former must be significantly larger than the latter, satisfying the condition in Eq.~\eqref{eq:condition}. This justifies our assertion in Section IIIA of the main text that the large errors in the electrostatic energy derived from the RI densities of silicon and aluminium arise from an inaccurate description of the electron density near the nucleus.

\begin{table}[t]
\centering
\setlength{\tabcolsep}{6pt}
\begin{tabular}{c | c c c c c } 
 Dataset & $\bar{\epsilon}_{\rho}^{\text{RI}}$ (\%) & $\bar{\epsilon}^{\text{RI}}_{xc}$ &  $\bar{\epsilon}^{'\text{RI}}_{xc}$ & $\bar{\epsilon}^{\text{RI}}_{el}$ &  $\bar{\epsilon}^{'\text{RI}}_{el}$  \\
 \hline
 Al & 0.002 & 0.04 & 0.04 & 2.65 & 2.28 \\
 Si & 0.003 & 0.01 & 0.00 & 3.57 & 0.19 \\
 I$_\mathrm{h}$ Ice & 0.01 & 0.00 & 0.00 & 0.13 & 0.03 \\
\end{tabular}
\caption{For the auxiliary basis constructed the using the overlap metric for orthogonalisation, the average error in the approximate RI density ($\bar{\epsilon}_{\rho}^{\text{RI}}$), along with the average error in exchange-correlation and electrostatic energies derived from it ($\bar{\epsilon}^{\text{RI}}_{xc}$ and $\bar{\epsilon}^{\text{RI}}_{el}$). These errors are relative to the QM reference values. $\bar{\epsilon}^{'\text{RI}}_{xc}$ and $\bar{\epsilon}^{'\text{RI}}_{el}$ are the ``baselined'' average errors, which remain after the mean error has been subtracted from each energy; this indicates the remaining error after the systematic error has been removed. All energies are reported in meV per atom, and are presented in comparison to those in Table I of the main text.}
\label{tab:svs_energy_err_ri}
\end{table}

We found that this systematic error could be significantly reduced by using an ``alternative auxiliary basis'', which is obtained by using an overlap metric when performing the Gram-Schmidt orthogonalisation required to eliminate linear dependencies in the auxiliary basis, rather than the Coulomb metric used to construct the ``standard auxiliary basis'' in FHI-aims, which has been used until this point. The average errors in the resulting RI densities are shown in Table \ref{tab:svs_energy_err_ri}, along with the average errors in the exchange-correlation and electrostatic energies derived from the RI density. By comparison to Table I in the main text, it is clear that this ``alternative auxiliary basis'' provides a superior RI density for aluminium and silicon. The RI density of the ice structures is largely unchanged, since changing the metric used in the orthogonalisation made very little difference to the resulting auxiliary basis.

\begin{figure}[t]
\includegraphics[width=0.75\linewidth]{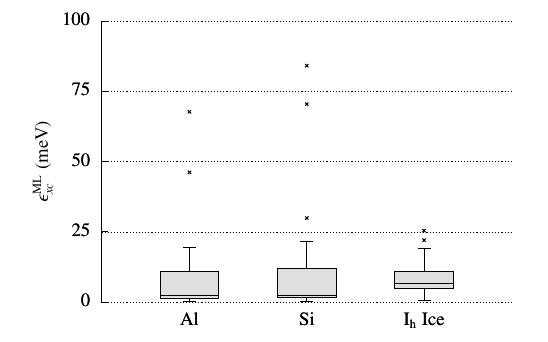}
\caption{For the auxiliary basis constructed the using the overlap metric for orthogonalisation, the distribution of the absolute errors in the exchange-correlation energy, $\bar{\epsilon}^{\text{ML}}_{xc}$, arising from the predicted density of 20 randomly selected structures from the three datasets.}
\label{fig:svs_xc}
\end{figure}

However, we found that this using this alternative basis led to significantly less stable results when using the SALTED method. As an example of this, Figure \ref{fig:svs_xc} shows the distribution of the absolute errors in the exchange-correlation energy derived from the density predicted by the SALTED method for 20 structures from the three datasets using this alternative basis. There are at least two significant outliers in both the aluminium and silicon datasets; this stands in contrast to the equivalent results obtained using the standard auxiliary basis in FHI-aims, shown in the middle panel of Figure 3 in the main text, which contain no outliers at all.

When using the standard auxiliary basis, the errors introduced to the electrostatic energies of aluminium and silicon by the RI approximation are smaller than those introduced by the SALTED method, and so we find that this is still an acceptable basis. Nonetheless, it is important to note that these systematic errors can be reduced by a suitable change to the auxiliary basis functions, but that functions which produced an better RI density are not necessarily better suited to the SALTED machine learning algorithm.

\section{SALTED hyper-paramaters for homogeneous datasets}

When training a SALTED model, there are three hyper-parameters which must be optimised. Two of them are associated with the spatial extent and resolution of the smooth Gaussian density that enters the $\lambda$-SOAP representation of the atomic structure, namely the radial cutoff $r_\text{c}$ of the atomic environment and the broadening $\sigma$ of the Gaussian functions. The third parameter is the regularization $\eta$, defined following Eq.~8 in the main text, which modulates the smoothness of the model and therefore its accuracy in an out-of-sample prediction. In addition, the model must be converged with respect to the number of atomic environments $M$ used in the sparse approximation of the density-coefficients in Eq.~7.

\begin{figure}
\includegraphics[width=0.6\linewidth]{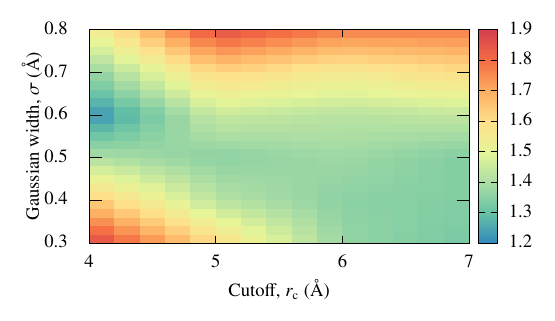}
\includegraphics[width=0.495\linewidth]{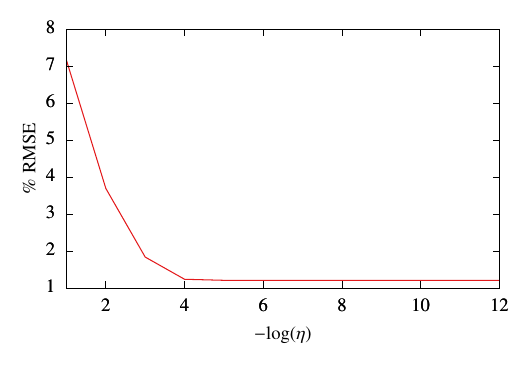}
\includegraphics[width=0.495\linewidth]{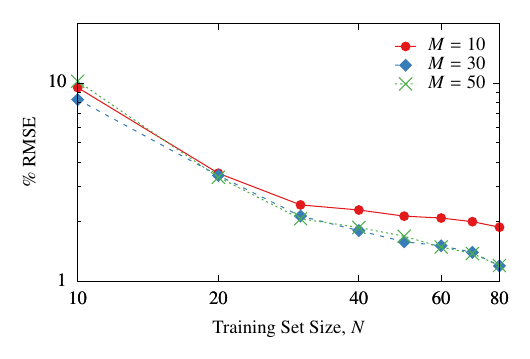}
\caption{The optimisation of the hyper-parameters for the aluminium dataset. The colourmap above shows the \% RMSE defined in Eq.~15 in the main text as a function of the two SOAP parameters $r_\text{c}$ and $\sigma$. The colourmap has been interpolated along each axis for clarity. Bottom left: the \% RMSE as a function of the regularization parameter $\eta$.  Bottom right: the \% RMSE as a function of the number of structures used in the training set $N$, using three different values of $M$, the number of atomic environments used in the sparse approximation to the coefficients.}
\label{fig:Al_optimisation}
\end{figure}

\begin{figure}
\includegraphics[width=0.6\linewidth]{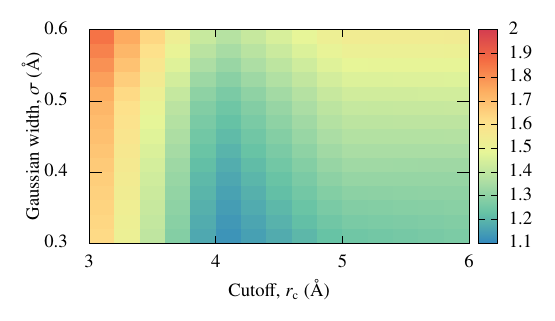}
\includegraphics[width=0.495\linewidth]{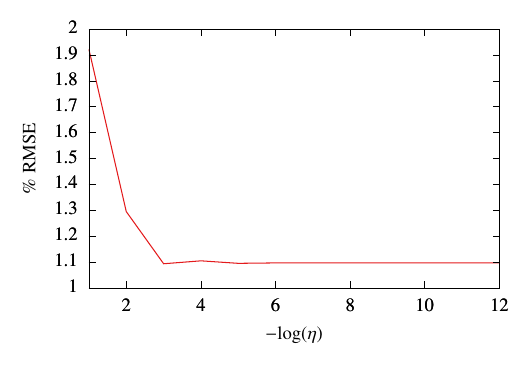}
\includegraphics[width=0.495\linewidth]{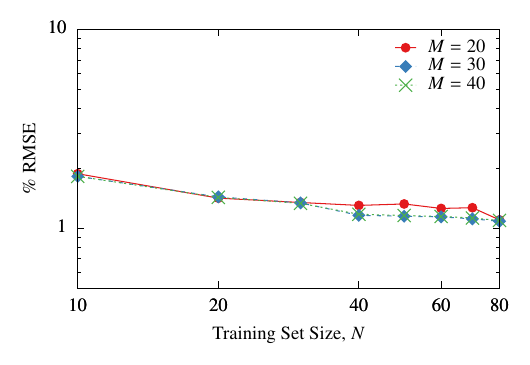}
\caption{The optimisation of the hyper-parameters for the silicon dataset. The colourmap above shows the \% RMSE defined in Eq.~15 in the main text as a function of the two SOAP parameters $r_\text{c}$ and $\sigma$. The colourmap has been interpolated along each axis for clarity. Bottom left: the \% RMSE as a function of the regularization parameter $\eta$.  Bottom right: the \% RMSE as a function of the number of structures used in the training set $N$, using three different values of $M$, the number of atomic environments used in the sparse approximation to the coefficients.}
\label{fig:Si_optimisation}
\end{figure}

\begin{figure}
\includegraphics[width=0.6\linewidth]{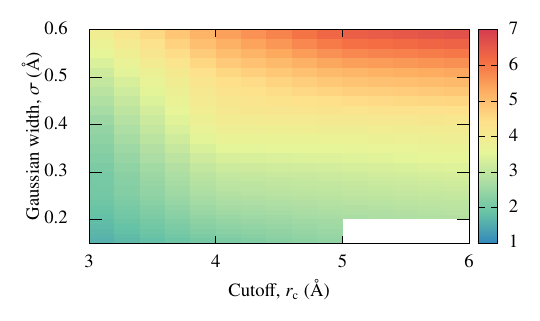}
\includegraphics[width=0.495\linewidth]{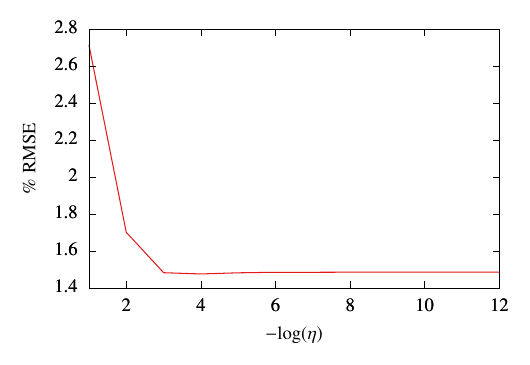}
\includegraphics[width=0.495\linewidth]{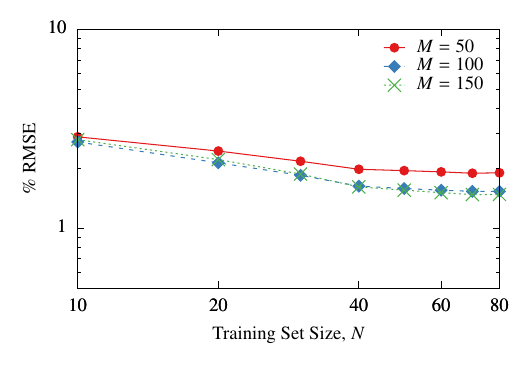}
\caption{The optimisation of the hyper-parameters for the ice dataset. The colourmap above shows the \% RMSE defined in Eq.~15 in the main text as a function of the two SOAP parameters $r_\text{c}$ and $\sigma$. The colourmap has been interpolated along each axis for clarity. Bottom left: the \% RMSE as a function of the regularization parameter $\eta$.  Bottom right: the \% RMSE as a function of the number of structures used in the training set $N$, using three different values of $M$, the number of atomic environments used in the sparse approximation to the coefficients. Using $r_\text{c}=6$ \AA~and $\sigma=0.15$ \AA~did not produce a stable regression model, so no \% RMSE is available for this combination of values. }
\label{fig:ice_optimisation}
\end{figure}

For each dataset, we first optimised the two SOAP parameters $r_\text{c}$ and $\sigma$ simultaneously, then the regularization parameter, and finally converged the learning curves with respect to $M$. The optimisation of the three hyperparameters was performed using a training set of 80 structures and a validation set of 20 structures. The results of this process are shown for the aluminium, silicon and ice datasets in Figs.~\ref{fig:Al_optimisation}, \ref{fig:Si_optimisation} and \ref{fig:ice_optimisation} respectively. The selected values of $r_{\text{c}}$, $\sigma$, $\eta$ and $M$ are given in Table \ref{tab:hyperparameters}.

The reference densities used in the main text are calculated with standard ``tight'' settings within AIMS, using converged k-point grids. For all aluminium structures, as well as the smaller silicon structures, a $(16 \times 16 \times 16)$ grid is used; for the larger silicon structures an $(8 \times 8 \times 8)$ grid is used. For the 4-molecule ice systems a $(4 \times 4 \times 4)$ grid is used, while the densities of the ice supercells are calculated at the Gamma point.

\begin{table}[b]
\centering
\begin{tabular}{c | c c c c } 
 Dataset & $r_{\text{c}}$ (\AA) & $\sigma$ (\AA) & $\eta$ & $M$ \\
 \hline
 Al & 4.0 & 0.6 & $10^{-4}$ & 50 \\
 Si & 4.0 & 0.3 & $10^{-5}$ & 40  \\
 I$_\mathrm{h}$ Ice & 3.0 & 0.15 & $10^{-4}$ & 150   \\
\end{tabular}
\caption{The selected hyperparameters for each dataset, along with the number of atomic environments required to converge the sparse approximation to the coefficients.}
\label{tab:hyperparameters}
\end{table}

\section{SALTED hyper-paramaters for heterogeneous datasets}

\begin{figure}
\includegraphics[width=0.6\linewidth]{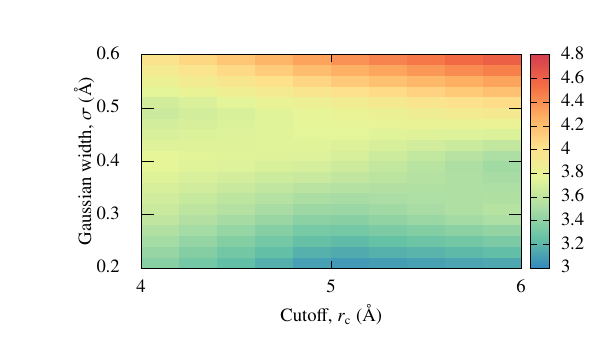}
\includegraphics[width=0.495\linewidth]{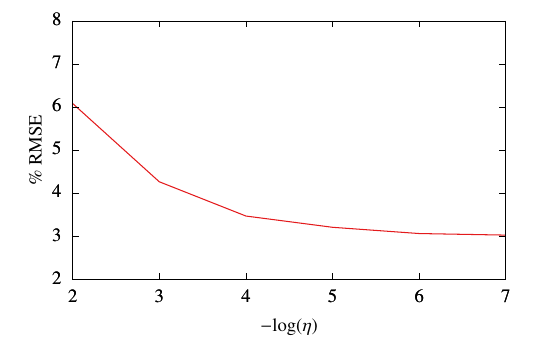}
\includegraphics[width=0.495\linewidth]{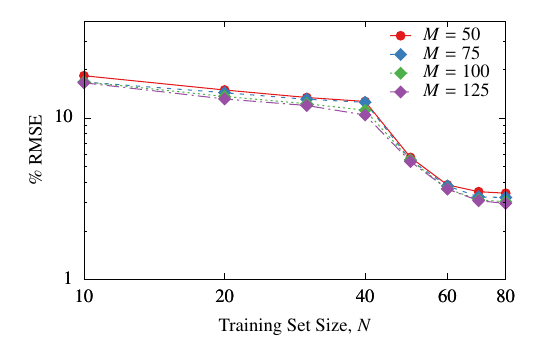}
\caption{The optimisation of the hyper-parameters for the mixed dataset comprised of the Al, Si and ice datasets. The colourmap above shows the \% RMSE defined in Eq.~15 in the main text as a function of the two SOAP parameters $r_\text{c}$ and $\sigma$. The colourmap has been interpolated along each axis for clarity. Bottom left: the \% RMSE as a function of the regularization parameter $\eta$.  Bottom right: the \% RMSE as a function of the number of structures used in the training set $N$, using three different values of $M$, the number of atomic environments used in the sparse approximation to the coefficients.}
\label{fig:full_optimisation}
\end{figure}

\begin{figure}
\includegraphics[width=0.6\linewidth]{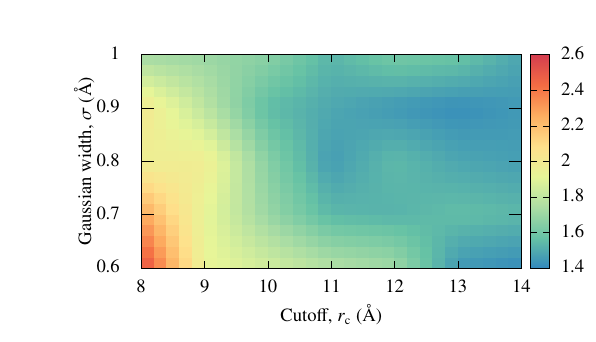}
\includegraphics[width=0.495\linewidth]{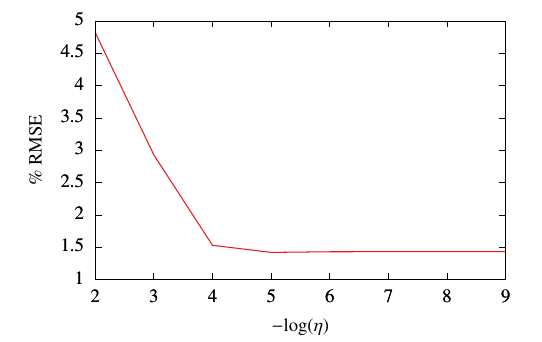}
\includegraphics[width=0.495\linewidth]{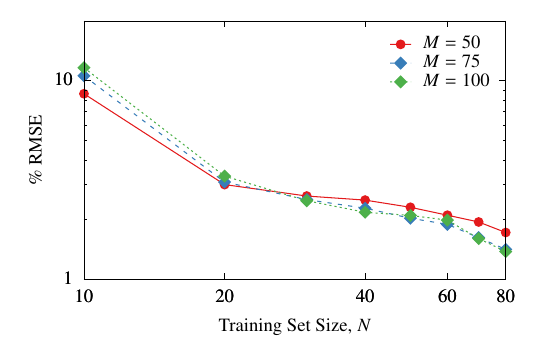}
\caption{The optimisation of the hyper-parameters for the perovskite dataset. The colourmap above shows the \% RMSE defined in Eq.~15 in the main text as a function of the two SOAP parameters $r_\text{c}$ and $\sigma$. The colourmap has been interpolated along each axis for clarity. Bottom left: the \% RMSE as a function of the regularization parameter $\eta$.  Bottom right: the \% RMSE as a function of the number of structures used in the training set $N$, using three different values of $M$, the number of atomic environments used in the sparse approximation to the coefficients. }
\label{fig:per_optimisation}
\end{figure}

To demonstrate the accuracy of SALTED when applied to heterogeneous datasets, first constructed a single dataset by combined the Al, Si and ice datasets, and re-optimised all of the learning parameters for this new set of structures. The results of this process are shown in Fig.~\ref{fig:full_optimisation}, with final hyperparameters of $r_\text{c} = 5$ \AA, $\sigma = 0.2$ \AA, and $\eta = 10^{-7}$. In addition, we applied SALTED to dataset of hybrid organic-inorganic perovskites, each containing one Sn and three F atoms, along with a small organic molecule. A $(4 \times 4 \times 4)$ $k$-grid was used to calculated the reference densities for these structures. The optimisation of the SALTED hyperparameters of this process are shown in Fig.~\ref{fig:per_optimisation}, with final hyperparameters of $r_\text{c} = 13$ \AA, $\sigma = 0.9$ \AA, and $\eta = 10^{-5}$ selected. Note that these values of $r_\text{c}$ and $\sigma$ are much larger than those obtained for the the other datasets due to the presence of the heavy Sn atom in the structures.

\section{Direct GPR hyper-parameters and learning curves}

\begin{figure}
\includegraphics[width=0.495\linewidth]{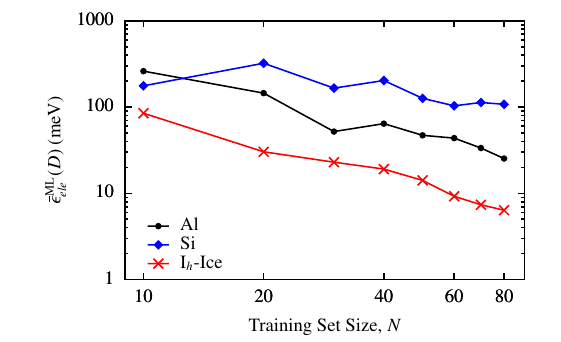}
\includegraphics[width=0.495\linewidth]{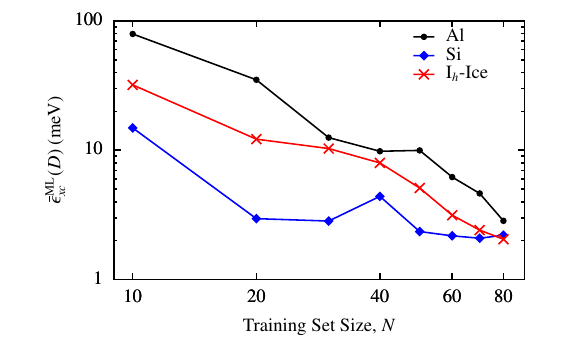}
\caption{Left: the direct learning curves for the predicted electrostatic energies of the 20 Al, Si and ice structures used to test the accuracy of the indirect energy predictions. Right: As the left plot, for the exchange-correlation energies.}
\label{fig:direct_lcs}
\end{figure}

The optimal values of the hyperparameters used in the direct GPR predictions of the electrostatic and exchange-correlation energies are provided in Table \ref{tab:direct_hyperparameters}. The three hyperparameters were optimised simultaneously, again using 80 training structures and 20 validation structures. The reported hyperparameters for ice were also used for the direct GPR predictions of the electrostatic and exchange-correlation energies of the ice supercells. The learning curves resulting from these hyperparameters are shown in Fig.~\ref{fig:direct_lcs}. Due to the small number of datapoints, these are not all perfectly monotonically decreasing. Nevertheless, in general they show satisfactory behaviour as the number of training points increases, with the exception of the electrostatic energy of Si, as noted in the main text.

\begin{table}[t]
\centering
\begin{tabular}{c | c c c } 
 Dataset & $r_{\text{c}}$ (\AA) & $\sigma$ (\AA) & $\eta$ \\
 \hline
 Al - Exchange-Correlation Energies & 4.0 & 0.5 & $10^{-8}$ \\
 Al - Electrostatic Energies & 4.0 & 0.4 & $10^{-5}$ \\
 Si - Exchange-Correlation Energies & 5.0 & 0.3 & $10^{-8}$  \\
 Si - Electrostatic Energies & 5.0 & 0.2 & $10^{-8}$ \\
 I$_\mathrm{h}$ Ice - Exchange-Correlation Energies & 3.0 & 0.3 & $10^{-5}$   \\
 I$_\mathrm{h}$ Ice - Electrostatic Energies & 3.0 & 0.3 & $10^{-5}$  \\
\end{tabular}
\caption{The selected hyperparameters for each dataset and property for the direct GPR predictions. In this case the regularisation parameter $\eta$ simply multiplies an identity matrix.}
\label{tab:direct_hyperparameters}
\end{table}

\begin{figure}[t!]
\centering
\includegraphics[width=0.6\linewidth]{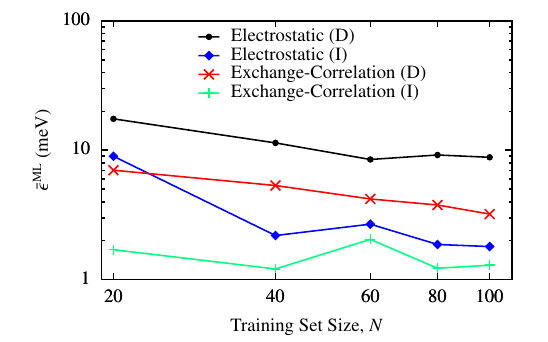}
\includegraphics[width=0.6\linewidth]{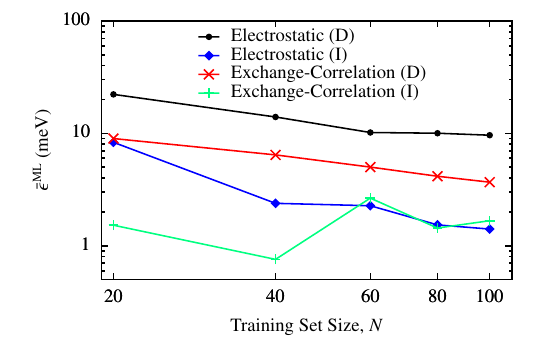}
\includegraphics[width=0.6\linewidth]{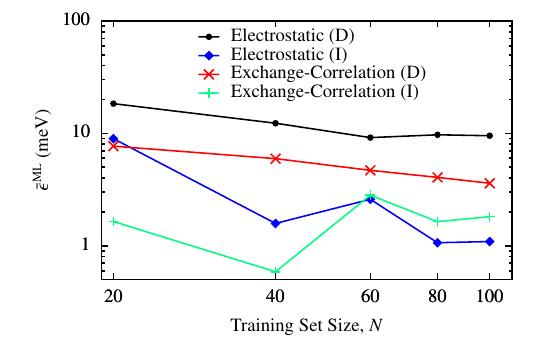}
\caption{Learning curves for the mean absolute errors in the exchange-correlation and electrostatic energies ($\bar{\epsilon}^{\text{ML}}_{xc}$ and $\bar{\epsilon}^{\text{ML}}_{el}$) derived from the predicted electron densities (the indirect errors, I), and predicted directly using Gaussian process regression (D) for the 128- (top), 256- (middle), and 512-molecule (bottom) ice supercell. These errors are relative to the QM reference values.}
\label{fig:direct_extrap_learning_curves}
\end{figure}

The learning curves for the exchange-correlation and electrostatic energies derived from the predicted electron densities (the indirect errors, I), and predicted directly using Gaussian process regression (D) for the 128-, 256- and 512-molecule ice supercell are shown in Fig.~\ref{fig:direct_extrap_learning_curves}. These all show qualitatively similar behaviour to Fig.~6 in the main text: the direct learning curves decrease monotonically, while the indirect learning curves are noisier, but across the learning curve the indirect method shows superior performance.

\section{Isolated molecules}

\begin{figure}
\includegraphics[width=0.75\linewidth]{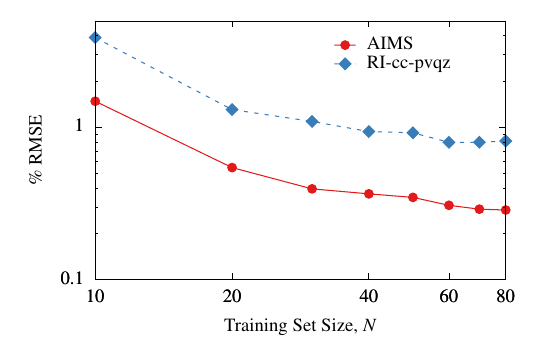}
\caption{The \% RMSE in the density of a test set of 200 water molecules, as a function of the number of structures used in the training set $N$, for which the density has been expanded using the numerical atom-centred orbitals as described in the main text.}
\label{fig:isolated}
\end{figure}

As is mentioned in the main text, the formalism presented here may be applied equally to periodic systems and isolated molecules. As an illustration of this, we show in Figure \ref{fig:isolated} the learning curve for a set of isolated water molecules, obtained using $r_\text{c} = 4$ \AA, $\sigma = 0.3$ \AA, $\eta = 10^{-5}$ and $M=100$. The test set consists of 200 configurations randomly selected from the full set of 1000 structures. This dataset has previously been used when assessing the accuracy of symmetry-adapted machine learning of tensors (PRL \textbf{120}, 036002, \emph{2018}). The learning curve decreases monotonically with the number of training structures, as expected, arriving at an error of approximately 0.3\% using 80 training structures, consistent with the results for the periodic examples shown in the main text. The average integrated mean absolute error in the density at this point is 0.23\%, which is consistent with errors found for periodic systems, and compares favourably to previous predictions of the density of isolated molecules.

Figure \ref{fig:isolated} also shows the learning curve obtained for the same structures, but using a Gaussian basis set to both calculate the QM reference density (specifically the cc-pvdz basis) and to represent the RI and ML densities (using the RI-cc-pvdz basis). The optimal hyper-parameters are found to be the same in both cases. The learning curve is broadly similar to that obtained using the numerical atom-centred orbitals of FHI-aims, although slightly less accurate.